\documentclass[times, twocolumn]{aastex62}
\usepackage{graphicx,amsmath,natbib}
\usepackage{color}
\usepackage{caption}
\usepackage{subcaption}

\newcommand{\Aerodyne}{Aerodyne Industries, 8910 Astronaut Blvd \#208, Cape Canaveral, FL 32920, USA}
\newcommand{\ATA}{ATA Aerospace, 7474 Greenway Center Drive, Greenbelt, MD 20770, USA}
\newcommand{\AURA}{Association of Universities for Research in Astronomy, 1331 Pennsylvania Avenue NW Suite 1475, Washington, DC 20004, USA}
\newcommand{\Ball}{Ball Aerospace and Technologies Corporation, 1600 Commerce Street, Boulder, CO 80301, USA}
\newcommand{\Genesis}{Genesis Engineering Solutions, Inc., 4501 Boston Way, Lanham, MD 20706, USA}
\newcommand{\Goddard}{NASA Goddard Space Flight Center, 8800 Greenbelt Road, Greenbelt, MD 20771, USA}
\newcommand{\Intuitive}{3700 Bay Area Blvd \#600, Houston, TX 77058, USA}
\newcommand{\JPL}{Jet Propulsion Laboratory, California Institute of Technology, 4800 Oak Grove Dr, Pasadena, CA 91109, USA}
\newcommand{\JSC}{NASA Johnson Space Center, 2101 E NASA Parkway, Houston, TX 77058, USA}
\newcommand{\KBR}{KBR, 6200 Guardian Gateway, Suite 105, Aberdeen, MD 21005, USA}
\newcommand{\Harris}{L3Harris Corporation, Space and Intelligence Systems, 400 Initiative Drive, P.O. Box 60488, Rochester, NY, USA 14606, USA}
\newcommand{\LambdaLLC}{Lambda Consulting LLC/Advanced NanoPhotonics, 4437 Windsor Farm Road, Harwood, MD 20776, USA}
\newcommand{\LASP}{Laboratory for Atmospheric and Space Physics, 1234 Innovation Dr, Boulder, CO 80303, USA}
\newcommand{\MSFC}{NASA Marshall Space Flight Center, 4600 Rideout Road, Huntsville, Alabama 35812, USA}
\newcommand{\NASAHQ}{NASA Headquarters, 300 E Street SW, Washington, DC 20546, USA}
\newcommand{\NGSS}{Northrop Grumman Space Systems, 1 Space Park Boulevard, Redondo Beach, CA 90278, USA}
\newcommand{\SAIC}{SAIC, 12010 Sunset Hills Road, Reston, VA 20190, USA}
\newcommand{\SAO}{Smithsonian Astrophysical Observatory, Cambridge, MA, USA}
\newcommand{\STScI}{Space Telescope Science Institute, 3700 San Martin Drive, Baltimore, MD 21218, USA}
\newcommand{\TMT}{Thirty Meter Telescope, 100 West Walnut Street, Suite 300, Pasadena, CA 91124, USA}
\newcommand{\Retired}{Retired}

\shorttitle{The Telescope for the James Webb Space Telescope Mission}
\shortauthors{McElwain, Feinberg, \& Perrin et al.}

\begin{document}

\title{The James Webb Space Telescope Mission:\\Optical Telescope Element Design, Development, and Performance}

% Primary authors
\author[0000-0003-0241-8956]{Michael W. McElwain}\affil{\Goddard}
\author{Lee D. Feinberg}\affil{\Goddard}
\author[0000-0002-3191-8151]{Marshall D. Perrin}\affil{\STScI}
\author{Mark Clampin}\affil{\NASAHQ}
\author{C. Matt Mountain}\affil{\AURA}
\author{Matthew D. Lallo}\affil{\STScI}
\author{Charles-Philippe Lajoie}\affil{\STScI}
\author{Randy A. Kimble}\affil{\Goddard}
\author{Charles W. Bowers}\affil{\Goddard}
\author{Christopher C. Stark}\affil{\Goddard}

% Alphabetical Authors
% ...

\begin{abstract}
The \textit{James Webb Space Telescope} (JWST) is a large, infrared space telescope that has recently started its science program which will enable breakthroughs in astrophysics and planetary science. Notably, JWST will provide the very first observations of the earliest luminous objects in the Universe and start a new era of exoplanet atmospheric characterization. This transformative science is enabled by a 6.6 m telescope that is passively cooled with a 5-layer sunshield. The primary mirror is comprised of 18 controllable, low areal density hexagonal segments, that were aligned and phased relative to each other in orbit using innovative image-based wavefront sensing and control algorithms.  This revolutionary telescope took more than two decades to develop with a widely distributed team across engineering disciplines. We present an overview of the telescope requirements, architecture, development, superb on-orbit performance, and lessons learned. JWST successfully demonstrates a segmented aperture space telescope and establishes a path to building even larger space telescopes.\\    
\end{abstract}

%\keywords{instrumentation: space telescope, segmented primary mirror, wavefront control}

\section{Introduction}

In the mid-1990s, the \textit{Hubble Space Telescope} (Hubble) observed the now-iconic Hubble Deep Field, which provided direct evidence that the universe evolved from the hot plasma left following the Big Bang to the galaxies and stars we see nearby. These observations revolutionized our understanding of our place in the universe. However, due to the intrinsic faintness of very distant galaxies and their wavelength shift into the infrared, observations of the very early universe remained beyond the capabilities of Hubble. Observing the first stars and galaxies required infrared wavelength coverage and increased sensitivity. This motivated the science community to conceive of the \textit{James Webb Space Telescope} (JWST), a large, general-purpose infrared-optimized observatory \citep{Dressler+1996} (Gardner PASP\footnote{Citations followed by PASP indicate papers to be included in the JWST Overview Special Issue.}). The early science motivations for JWST were organized into four broad themes: origins of the universe (including first light and reionization), the assembly of galaxies, the birth of stars and protoplanetary systems, and the formation of planetary systems and the origins of life. The 2000 Astrophysics Decadal Survey, ``Astronomy and Astrophysics in the New Millenium'' placed the JWST concept as the top priority for NASA \citep{NAS-Astro2000}. An Independent Review Board in 2018 and the recent 2020 Astrophysics Decadal survey concluded the mission was as compelling two decades later as when it was originally conceived \citep{NAS-Astro2020}. 

Each of the four science themes demanded a large ($\sim$25~m$^{2}$), infrared-space telescope covering a wavelength range from 0.6 to 28.1 $\mu$m \citep{Gardner+Mather+Clampin+etal_2006}. The paradigm for space telescope architectures needed to change in order to make this space telescope with a 6.6~m circumscribed diameter primary mirror, as this physical size is larger than the 5.4~m rocket fairing diameters available at the time. The telescope and observatory needed to be stowed for launch and then undergo a series of major deployments to transform it into the operational configuration. The solution was to make the primary mirror segmented based on implementations from the W.~M. Keck Observatory's (Keck) two ground-based telescopes. The JWST primary mirror is comprised of 18 hexagonal segments, each of which is supported by a common backplane opto-mechanical structure and adjustable in 6 positional degrees of freedom, as well as radius of curvature. The primary mirror segments assemblies, large precision cryogenic structures, and the wavefront sensing and control did not exist and needed to be created specifically for JWST. The infrared observations means that the entire telescope and instruments must operate at cryogenic temperatures, which are reached through passive cooling by a 5-layer sunshield.

The JWST Yardstick study established an architecture that demonstrated the science performance and technical feasibility \citep{Bely_1998}, which baselined passive cooling via the large sunshield while operating in orbit at L2. Early architecture concepts were competed with a proposal down-select that was awarded to TRW Inc., now Northrop Grumman Corporation (Northrop), and Ball Aerospace, along with a team at ATK, now Northrop, and Kodak, now L3Harris Technologies. The architecture concept studies further refined the design \citep{Lightsey+Ebbets_2000}. During the concept studies, the JWST primary mirror was 8~m in diameter with multiple mirror concepts. Once TRW Inc. was selected in 2002, their initial design was based on a 36 segment architecture that was 7~m in diameter.  However, each segment had only tip, tilt, piston and radius of curvature control.  Later a trade was done to consider an 18 segment design that while only 6.6~m in diameter would include seven degrees of freedom on each mirror using a hexapod with a radius of curvature actuator. A key consideration for the trade was that 18 segments reduced the amount of segment perimeters (edges) which are one of the key challenges of making mirrors.  Ultimately the decision was to baseline 18 segments and hexapods, which enabled astigmatism correction that was beneficial in the mirror manufacturing, alignment tolerances, and cryogenic testing program.

The JWST observatory architecture, science instrument on-sky performance, backgrounds, and science performance are presented in this PASP JWST Overview Special Issue. In this paper, we focus on the telescope. We start by describing the driving requirements in Section~\ref{sec:TelescopeRequirements}, and the resulting design architecture in Section~\ref{sec:TelescopeArchitecture}. Highlights from the development and integration and test phases are presented in Section~\ref{sec:TelescopeDevelopment}. We present the commissioning and on-orbit performance of the telescope in Section~\ref{sec:OnOrbitCharacterization}. In Section~\ref{sec:ScienceEra}, we report on the science era characterization. A discussion of lessons learned is presented in Section~\ref{sec:LessonsLearned}. We conclude in Section~\ref{sec:conclusion}.

The on-orbit performance of the telescope is better than the requirements of the telescope across the board \citep{Rigby+Perrin+McElwain+etal_2022}, a result which was achieved through diligent systems engineering and a thorough test program. JWST's success demonstrates that ambitious space telescopes can be built to help answer some of the biggest questions facing humanity, such as how the universe works and how we got here.

\section{Telescope Requirements}
\label{sec:TelescopeRequirements}

The telescope architecture was designed to meet just a few driving optical requirements derived by early science trade studies, as presented in Table~\ref{tab:RequirementSpace} alongside the on-orbit performance values. A large, primary mirror aperture was required to have a total unobscured collecting area greater than 25~m$^{2}$ that operated over the spectral range 0.6~$\mu$m to 27~$\mu$m. The optical area x transmission product was the metric used to specify the performance of the coatings, contamination, and micrometeoroid damage. For this large aperture, the telescope image quality was specified using the Strehl ratio metric at 2~$\mu$m\footnote{A key requirements downscope was dropping the formal requirement for 1~$\mu$m performance, to reduce complexity in the integration and test (I\&T) program, with the expectation that an optical system meeting or exceeding requirements at 2~$\mu$m would necessarily also yield very good performance at 1$\mu$m. This has proven the case in flight. See Section~\ref{subsec:VerSciPerformance} for lesson learned.} and 5.6~$\mu$m, with a diffraction-limited image quality value of 0.8 over the NIRCam and MIRI fields of view, respectively. Finally, a set of encircled energy stability requirements were defined at time intervals of 24 hours and 14 days, specifying less than 2.3~$\%$ and 3.0~$\%$ RMS variations of the energy within a 80~mas radius aperture at 2$\mu$m wavelength following a worst case hot-to-cold slew within the field of regard (FOR).

% MASTER TABLE INCLUDES Optical Requirements, Ground Predictions, On-Sky Performance. See PLAR \S3.7 Chart 3.\\
\begin{deluxetable}{ccc}
\tabletypesize{\scriptsize}
\tablecaption{Optical Requirements \label{tab:RequirementSpace}}
\tablewidth{0pt}
\tablehead{ 
\colhead{Optical Requirements} & 
\colhead{Requirement} & 
\colhead{Measured}
}
\startdata
Optical Area in $m^{2}$ & $>$25.0  & 25.44 \\ 
Strehl ratio over NIRCam FOV at 2.0~$\mu$m & $>$0.8  & 0.84 \\ 
Strehl ratio over MIRI FOV at 5.6~$\mu$m & $>$0.8  & 0.92 \\ 
Encircled Energy Stability over 24 hours\tablenotemark{a} & $<$2$\%$  &  0.2$\%$\tablenotemark{c} \\ 
Encircled Energy Stability over 14 hours\tablenotemark{b} & $<$3$\%$ & 0.53$\%$\tablenotemark{c} \\ 
Vignetting & None  & None \\ 
%Wavelength Coverage from 0.6-27~$\mu$m & & gold-coated mirrors \\ 
\enddata
\tablenotetext{a}{Aperture radius of 80~mas at a wavelength of 2$\mu$m from the mean over a 24 hour period, evaluated for a worst-case 10\degr pitch maneuver.}
\tablenotetext{b}{Aperture radius of 80~mas at a wavelength of 2$\mu$m from the mean over a 14 day period, evaluated for a worst-case hot-to-cold pitch maneuver.}
\tablenotetext{c}{Typical performance values. Occasional larger instabilities are observed due to tilt events, discussed in Section~\ref{subsubsec:TiltEvents}.}
\end{deluxetable}

The radiometric sensitivity for the observatory required the observatory to have low backgrounds that enabled the detection of faint sources. The largest near-infrared background component was to be from zodiacal light. In the longer wavelengths of the mid-infrared channel, the requirements permitted the background to be dominated by thermal self-emission from the observatory rather than the natural sky, but with that thermal emission tightly constrained to yield the required unprecedented mid-infrared sensitivity. The near-infrared background was limited by the observatory design and modeling, careful control of particulate levels, and the use of low reflectivity baffle materials. The low-background, mid-infrared environment is enabled by passively cooling the telescope and science instruments to cryogenic operating temperatures for the life of the mission. The observatory operates in a Lissajous orbit about the Earth-Sun L2 point that permits a sunshield to continuously keep the telescope and science instrument protected from the sun. (Rigby PASP Backgrounds) provides a detailed discussion of the backgrounds and the on-sky measurements.

All of the performance requirements were defined to be met at 5.5~yr after launch, referred to as ``End of Life'' (EOL), as that was the minimum required lifetime of the mission\footnote{The only consumable onboard JWST is the propellant, which is predicted to have a lifetime of greater than 20 years (Menzel PASP Observatory).}. The optical performance properties are expected to degrade over time due exposure within the space environment (e.g., Section~\ref{subsubsec:micrometeoroids}). The optical performance requirement predictions accounted for these degradations and used the worst-case prediction for each input category. For example, the encircled energy stability requirement used the worst-case wavefront error, thermal distortion, and image motion all happening at the same time, which is a condition that is known to be statistically unlikely to occur. The worst case predictions also included model uncertainty factors (MUFs) to manage uncertainties and provide margin at the system level. Therefore, the performance measured during commissioning, the so-called ``Beginning of Life'' (BOL) optical properties, should not be directly compared to the requirement EOL values.

In addition to the driving performance requirements, the telescope had challenging design constraints in order for the observatory to meet its system-level requirements. The telescope total mass was specified to be less than 2460~kg and the measured mass at launch was 2339kg.  The center of mass lateral location was specified to fit within a 200mm$\times$200mm envelope and the Observatory fit within that envelope with 37.9~mm margin to the nearest boundary. The stowed telescope volume was set to be within an envelope 3.985~m$~\times~$4.114~m~$\times$~6.942~m (V1,V2,V3). Both the mass and stowed telescope volume requirements were driven by the capabilities of the Ariane 5 launch vehicle. The deployed telescope volume expanded considerably to 9.386~m~$\times$~6.100~m~$\times$~7.971~m, which needed to be shaded by the sunshield while pointing within the designated field of regard. Finally, the power consumption of the telescope was set to be less than 50~W (measured 34~W) such that the total power budget for the observatory was maintained.

% Actual OTE mass was 2333 kg. and power was 33 W.

% Observatory Architecture
\begin{figure}
  \centering
%  \vspace{-0.7in}
  %\hspace{-1.5in}
  \includegraphics[width=0.45\textwidth]{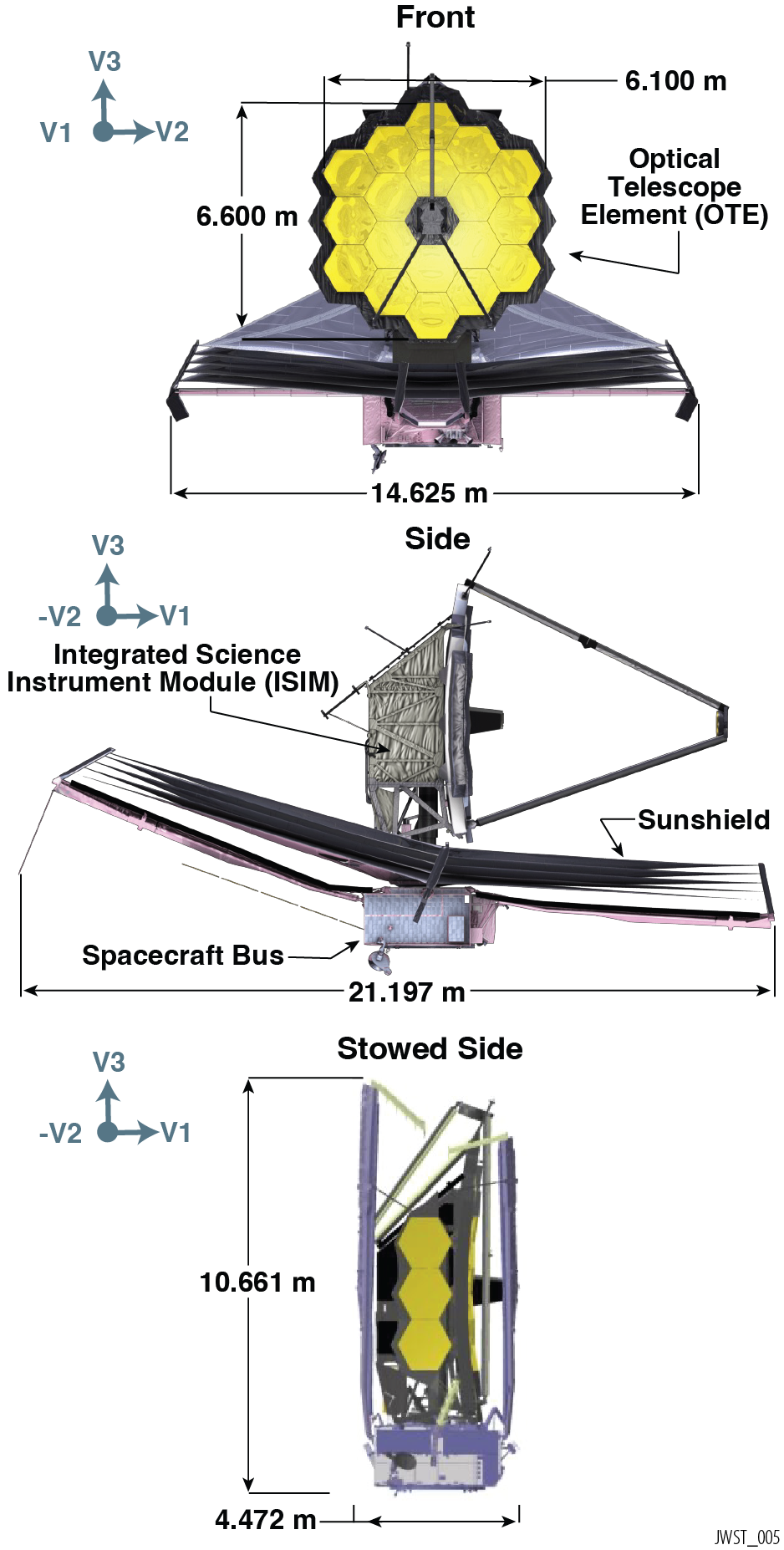}
%   \vspace{-0.9in}
  \caption{The telescope and science instruments are separated from the spacecraft bus and shielded from direct sunlight by a 5-layer sunshield. The telescope's primary and secondary mirrors are open to the celestial sky. When stowed (below), the observatory volume is significantly reduced to fit within the Ariane 5 fairing and pinned mechanisms enable the structure to withstand the launch environment. The V coordinate system origin is at the vertex of the primary mirror surface, along the boresight.} 
  \label{fig:ObservatoryArchitecture}
\end{figure}

\section{Telescope Architecture}
\label{sec:TelescopeArchitecture}

\begin{figure*}
  \centering
%  \vspace{-0.7in}
  %\hspace{-1.5in}
  \includegraphics[width=0.75\textwidth]{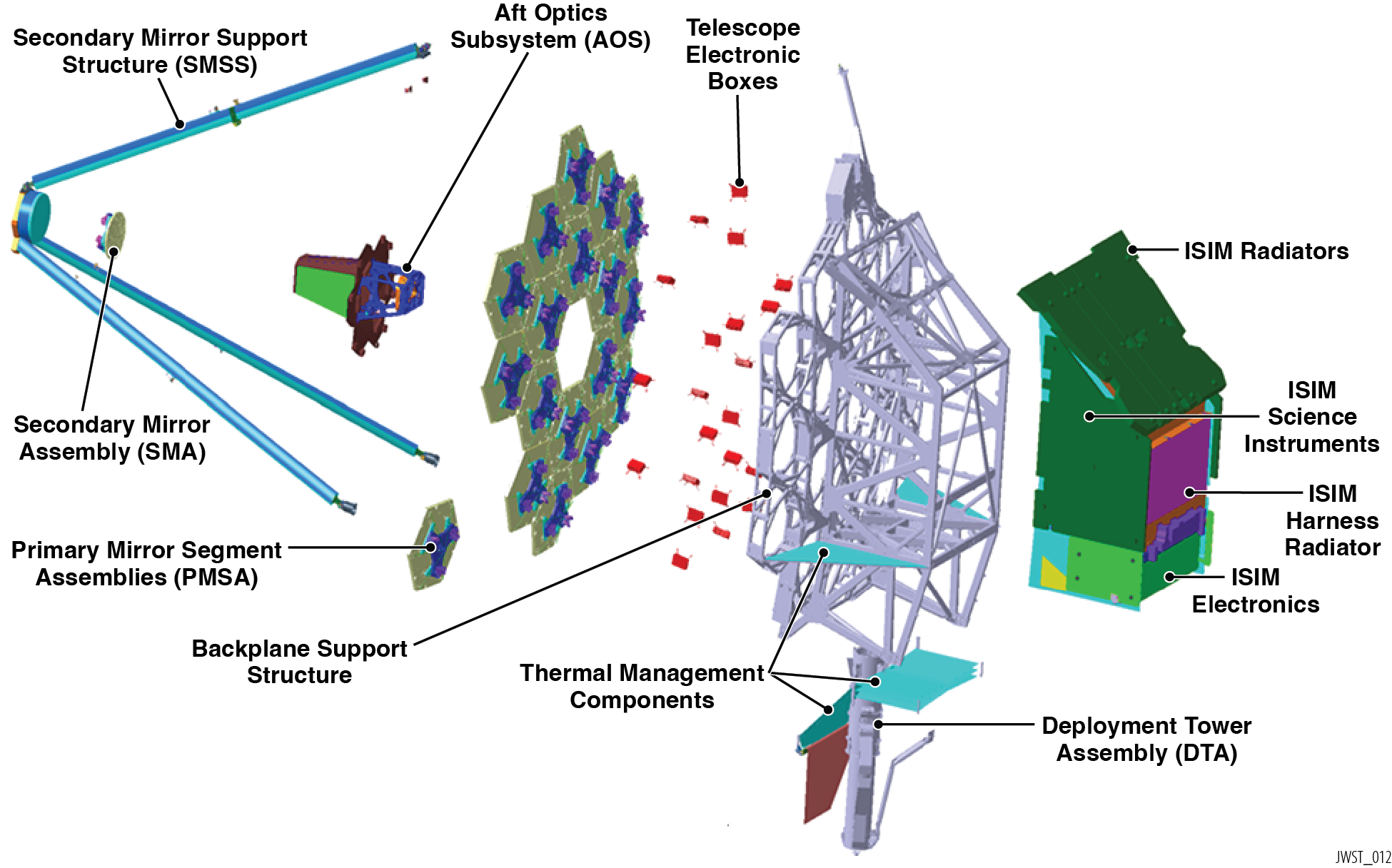}
%   \vspace{-0.9in}
  \caption{This exploded view shows telescope components, including the telescope mirrors, the optomechanical structures, control electronics, and the thermal management system. The integrated science instrument module includes the four science instruments.} 
  \label{fig:TelescopeExplodedView}
\end{figure*}

The observatory architecture consists of three major systems: the telescope and science instruments, the 5-layer sunshield, and the spacecraft bus \citep{Nella+Atcheson+Atkinson+etal_2004, Lightsey+Atkinson+Clampin+etal_2012}. Renderings of the observatory in its operational and stowed configurations are shown in Figure~\ref{fig:ObservatoryArchitecture}. The telescope and the science instruments are passively cooled by the sunshield and thermally isolated from the spacecraft bus and solar array that are on the warm, sun-facing side of the observatory. The telescope's primary mirror (PM) and secondary mirror (SM) are open to the celestial sky, but the tertiary and fine steering mirror are enclosed within the aft optics system (AOS). The telescope Cassegrain focus is formed near the entrance to the AOS where an aperture mask is located for stray light suppression. There are four science instruments (SIs): the NIRCam \citep{Rieke+Kelly+Horner_2005}, NIRSpec \citep{Jakobsen+Ferruit+AlvesDeOliveira+etal_2022}, MIRI \citep{Rieke+Wright+Boker+etal_2015,Wright+Wright+Goodson+etal_2015}, and FGS/NIRISS \citep{Doyon+Hutchings+Beaulieu+etal_2012}. All of the science instruments are mounted within the same structure, called the integrated science instrument module (ISIM), which is blanketed from the celestial sky. The instrument drive electronics and radiators are mounted on the exterior of the ISIM. Figure~\ref{fig:TelescopeExplodedView} shows an exploded view of the telescope components along with the science instruments, thermal management system and mechanical supports.   

The telescope field of regard is restricted to pointings that shield the telescope and science instruments from direct solar illumination. The telescope boresight pitch constraints are between pitch angles of 85$\degr$ and 130$\degr$ (0$\degr$ pointed towards the sun), roll about the telescope boresight is constrained to $\pm$5$\degr$, and yaw is unconstrained to a full 360$\degr$ around the sunline. The observatory is in an orbit around the Earth-Sun L2 Lagrange libration point, such that it orbits the sun along with the Earth over one year. While the Earth and L2 Lagrange point sweep along the celestial sphere, the field of regard on-sky visibility changes. At any instantaneous moment, 40\% of the sky is visible to the telescope, while over the course of the year the full sky is observable. The total visibility period over the course of a year increases with ecliptic latitude, ranging from $\sim$100 days in the ecliptic plane to continuous visibility in 5-degree-radius cones at the ecliptic poles.

The JWST telescope was designed to be stowed in order to fit within the Ariane 5 fairing for launch. Following launch, the observatory needed to be deployed into the operational configuration. The 18 primary mirror segments are arranged in a close-packed configuration with twelve segments in the center section and three segments on each of two wings which were folded back for launch. No segment is located at the nominal center position, as it is filled with the aft optical structure that holds the telescope tertiary and fine steering mirrors. The secondary mirror was also stowed such that it was face down above the center section, shielding its surface from particulate redistribution during launch. The telescope deployments included separating from the spacecraft bus using a deployable tower assembly, driving the secondary mirror into position, rotating and latching the primary mirror wings into position, and finally deploying the primary mirror segment assemblies and the secondary mirror from their launch locks. The telescope alignment used image-based sensing and control with the primary mirror segments and secondary mirror.

\subsection{Design Implementation} \label{subsec:DesignImplementation}

\begin{figure}
  \centering
%  \vspace{-0.7in}
  %\hspace{-1.5in}
  \includegraphics[width=0.5\textwidth]{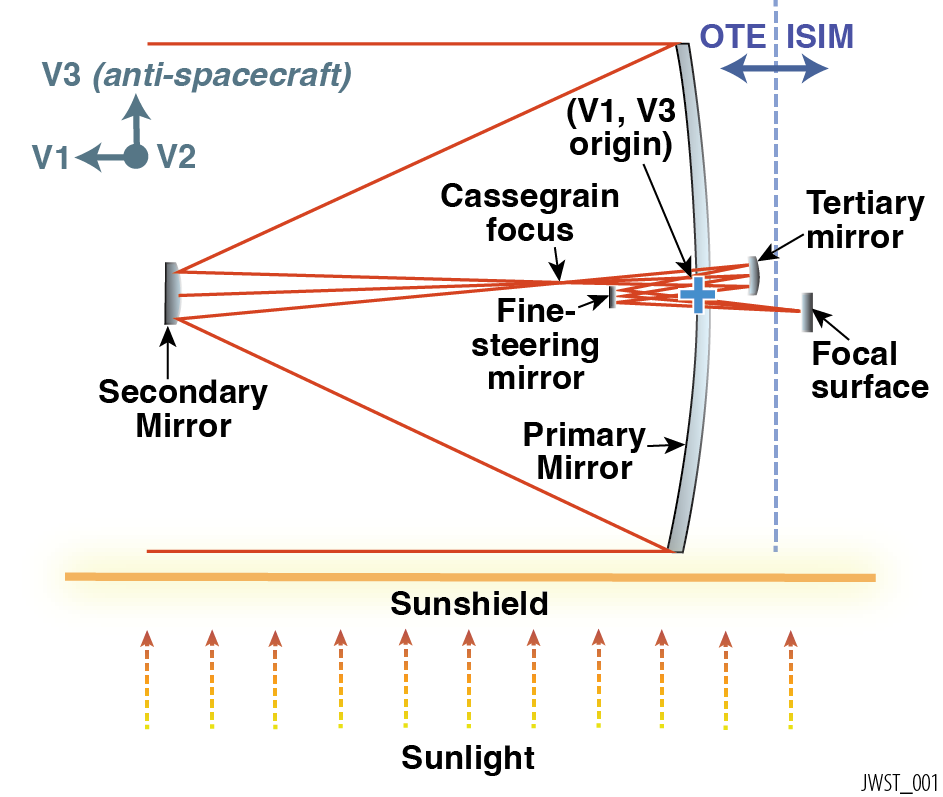}
%   \vspace{-0.9in}
  \caption{The JWST telescope is a three mirror anastigmat that has a primary mirror, secondary mirror, tertiary mirror, and a fine steering mirror. Each of the four science instruments has one or more pickoff mirrors near the telescope's focal surface such that they see different fields of view.} 
  \label{fig:OpticalDesign}
\end{figure}

\begin{figure}
  \centering
%  \vspace{-0.7in}
  %\hspace{-1.5in}
  \includegraphics[width=0.32\textwidth]{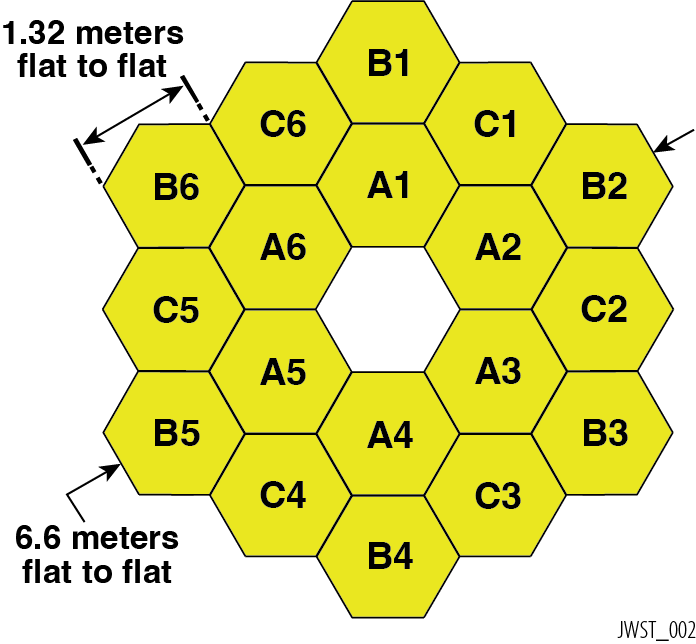}
%   \vspace{-0.9in}
  \caption{The primary mirror dimensions and the tiling configuration with 18  hexagonal-shaped segments to form a tricontagon. The A-, B-, and C-segment prescriptions are separate and take advantage of the six-fold symmetry.} 
  \label{fig:PrimaryMirror}
\end{figure}

The optical design of the telescope is a three-mirror anastigmat (TMA) that corrects spherical aberration, coma, astigmatism, and field curvature (see \cite{Korsch_972}). In addition to the primary, secondary, and tertiary mirrors of the TMA, the telescope also includes a fine steering mirror (FSM) which is actively controlled to stabilize the line-of-sight optical path into the science instruments. A raytrace through the telescope with annotations for the optical surface is shown in Figure~\ref{fig:OpticalDesign}. Each of the 18 primary mirror segments are 1.32~m (flat-to-flat) separated by 7~mm gaps (see Figure~\ref{fig:PrimaryMirror}).  The PM circumscribed diameter is 6.64~m (segment flat-to-flat diameter is 6.575~m, entrance pupil diameter is 6.6~m) with an area equivalent to a 5.7~m circular filled aperture.  The telescope is f/20 with an effective focal length of 131.4~m. The field is about $\sim$ 18$\arcmin$ $\times$ 9$\arcmin$ at the telescope focal plane. Each of the four science instruments has one or more pick-off mirrors near the focal plane of the telescope such that they observe offset parts of the sky.  The telescope optical parameters are reported in Table~\ref{tab:OpticalParameters}.

% Optical Design

% Primary mirror

Each of the 18 primary mirror segments is controlled in the six mechanical degrees of freedom (DoF) using a hexapod with six actuators. The segments are semi-rigid with a radius of curvature actuator at the center of each. During the alignment process, the radius of curvature for each primary mirror segment is matched and the primary and secondary mirrors are optimally aligned. The gaps between segments, nominally 7~mm, were tracked during the deployment and alignments by bookkeeping the actuator resolver counts and independently monitoring the positions using linearly variable differential transformers (LVDT) electromechanical sensors. All of the telescope optics are made from Beryllium based primarily on its low coefficient of thermal expansion (CTE) over the mirror's operating temperature range of $\sim$35-55 K. Segments closer to the warm spacecraft core region have correspondingly higher temperatures.

Mirror moves are controlled through the actuator drive unit (ADU) that is housed on an electronics panel within the spacecraft bus. The ADU provides state-of-health telemetry for the telescope hardware, controls the 132 mirror actuators, and polls telemetry like motor revolutions and strain gauges on the telescope.  Those signals pass through cryogenic electronics boxes: first the cold multiplexer units (CMUs) located on the telescope backplane, and a cold junction box (CJB) before returning to the ADU. In a separate mode, the ADU controls the fine steering mirror linear voice coil motors and provides telemetry for its x,y position, current, and temperatures.

Passive stability is achieved through mechanical $\sim$1~Hz isolators between the telescope and the spacecraft bus and a thermally stable backplane support for the telescope optics made from a lightweight composite. The reaction wheel assemblies (RWA) and cryocooler compressor assemblies housed within the spacecraft include vibration attenuators within their subsystems. The composite truss structure is comprised of a center section that supports 12 segments, and two wing sections, each supporting 3 segments. There is a separate backplane support frame that carries the load from the telescope structure and the ISIM. The secondary mirror is mounted to a tripod that consists of four composite tubes and connects to the primary center section in a four-bar linkage (see Figure~\ref{fig:TelescopeExplodedView}).

% OPTICAL PARAMETERS TABLE\\
\begin{deluxetable}{cccccccc}
\tabletypesize{\scriptsize}
\tablecaption{Telescope Optical Parameters \label{tab:OpticalParameters}}
\tablewidth{0pt}
\tablehead{ 
\colhead{Mirror} & 
\colhead{RoC} & 
\colhead{Surface} &
\colhead{Conic} &
\colhead{V1} & 
\colhead{V2} & 
\colhead{V3} &
\colhead{Phys. Size} \vspace{-0.2cm}\\ 
& \colhead{(mm)} & & & \colhead{(mm)} & \colhead{(mm)} & \colhead{(mm)} & \colhead{(mm)}}
\startdata
Primary & 15879.7 & concave & -0.9967 & 0 & 0 & 0 & 6605.2\\
Secondary & 1778.9 & convex & -1.6598 & 7169.0 & 0 & 0 & 738\\
Tertiary & 3016.2 & concave & -0.6595 & -796.3 & 0 & -0.19 & 728$\times$517\\
Fine Steering &  & flat & & 1047.8 & 0 & -2.36 & 172.5\\
\enddata
%\vspace{-0.8cm}
\tablecomments{The primary-to-secondary effective focal length (EFL) is 59400 mm., and the three-mirror telescope EFL is 131400 mm.}
\end{deluxetable}

Active pointing stability is achieved using a control loop that senses line of sight pointing with a fine guidance sensor (FGS, $\sim$16~Hz) and corrects pointing errors using the FSM. This controls line of sight pointing drifts with a 0.78~Hz control loop bandwidth. In addition, as the fine guidance sensor can only provide pointing measurements about two axes (image tip and tilt), the coarse pointing roll controller uses star tracker and inertial reference unit (IRU) measurements to stabilize the roll (clocking) about the boresight. However, thermal distortions may cause drifts between the star-tracker-based spacecraft pointing frame and the line of sight frame, which cannot be measured and therefore would remain uncorrected. This could create a rotational drift about the guide star, which manifests as image smear elsewhere in the science instrument field of view (see Section~\ref{subsubsec:ThermalStability}). Additionally, higher frequency image motion, due to reaction wheel and cryocooler disturbances, results in a Gaussian image blur. The image motion has a negligible effect on the encircled energy, but may degrade the image quality defined by the Strehl ratio requirement.

The telescope open architecture for passive cooling makes it susceptible to stray light from the celestial sky and the observatory emission itself. There are successive layers of stray light protection used to shield the science instrument focal planes. The sunshield protects the optical path from direct sunlight and enables passive cooling of the telescope and science instruments that reduces the self emission. The spreader bars that hold the sunshield layers in position have baffles on their caps, called epaulettes, that shield the optical path from the warm spreader bars. 

There is an oversized, internal pupil stop near the FSM to transfer the maximum PM collected energy and provide a well defined pupil for wavefront sensing.  A lightweight structure, called the `frill' (see Figure~\ref{fig:ObservatoryArchitecture} top), extends from the perimeter of the PM and substantially fills in the gap between the oversized pupil stop and the image of the PM at the stop. By filling in this gap, the frill blocks celestial light from behind and around the PM (`truant' stray light path) from becoming a source of background stray light. A similar baffle, called the bib, extends below the frill and blocks the direct path to the warm spacecraft core area. 

The AOS enclosure blocks stray light around the tertiary and FSM mirrors. The AOS entrance aperture is near the Cassegrain focus of the telescope and provides an aperture stop for stray light suppression.  A baffle extending above the FSM combined with careful telescope alignment, prevents light from the sky passing through the AOS entrance aperture and striking any instrument optics directly. This stray light path, called the rogue path, was identified early in the design phase and precautions were taken to keep light from this path from propagating directly through the SI optical trains to their detectors. In flight, however, scatter paths off SI mechanical structure were found to produce unwanted backgrounds from sources in this small region of sky (see Section~5.3 in Rigby PASP Science, Section \ref{sec:opticalmodeling-lesson} lesson learned). Fortunately, the rogue path stay light can be largely managed through observation scheduling. 

The integrated science instrument module is enclosed and blanketed for stray light and thermal purposes. Each of the science instruments also has an enclosure and the optical paths include internal baffles.

\section{Telescope Development}
\label{sec:TelescopeDevelopment}

The telescope development was a long and complex process that took place between 2002 and 2022. In this section, we provide an overview of the telescope development phase with discussions of the systems engineering, the new technologies developed, and the integration and testing needed to verify the design and workmanship. 

Prime contractor Northrop Grumman was responsible for the telescope design and built a team that included optics lead Ball Aerospace, composite lead ATK (later to be bought by Northrop Grumman), and the L3Harris Corporation (originally part of Kodak, then ITT Exelis) for their large optics integration heritage. The lead government organization was NASA's Goddard Space Flight Center, which took full responsibility for the telescope and provided key facilities where the mirrors and science instruments were integrated with the backplane structure and tested at ambient temperatures for vibration and acoustics. The Jet Propulsion Laboratory and NASA's Marshall Space Flight Center helped with technology development and mirror testing, while cryogenic tests took place at NASA's Johnson Space Flight Center. The Space Telescope Science Institute provided leadership in developing telescope operations plans and helped lead the commissioning efforts. Throughout the development phase, groups of scientists and engineers worked collaboratively across these organizations to jointly design, integrate, test, and commission JWST.

Further, the telescope development team sought input from external expertise for technical decision making. The project formed an independent Product Integrity Team (PIT) to provide technical engineering advice on the telescope. The PIT was led by Professor Duncan Moore of University of Rochester and Professor James Wyant of University of Arizona and included space- and ground-telescope experts from Hubble, the \textit{Spitzer Space Telescope} (Spitzer), Keck and many others.  Throughout the development phase, the PIT independently reviewed all aspects of the optics program, especially the integration and test campaign with its direct ties to the verification plans. The early test concepts matured and improved significantly over time with input from the optical PIT (e.g., \cite{Feinberg+Hagopian+Diaz_2006}). In cases where performance predictions threatened requirements, the JWST science requirements advisory board was convened, with representation from the JWST Science Working Group, to evaluate the science impacts.

The telescope design process began with the mirror substrate and moved outwards to the backplane and ultimately to the whole telescope and observatory design. New technologies were developed in parallel. Sub-system development schedules were managed such that the telescope would meet the larger observatory schedules. Early engineering design units (EDUs) of the primary mirror segment and secondary mirror were important to prove out the manufacturing process for these critical-path components. The primary mirror EDU demonstrated technology readiness, TRL-6, by carrying out successful acoustic and vibration tests that were not completed in the earlier technology development period. An engineering design unit partial version of the OTE center section, called Pathfinder, allowed verification of assembly, handling, and testing techniques. 

\subsection{Systems Engineering for the JWST Telescope} 

The JWST telescope development relied on systems engineering principles, tools, and practices as described in Menzel PASP Observatory. The systems engineering approach for JWST is presented thoroughly in \cite{Lightsey+Arenberg_2018}. The telescope performance requirements relied on detailed budgets for optical performance \citep{Lightsey+Chaney+Gallagher+etal_2010}, alignments \citep{Glassman+Levi+Liepmann+etal_2016}, and actuator ranges \citep{Barto+Acton+Finley+etal_2012}. The budgets were used throughout the mission development: to make design decisions and carry out concept trades, to design the integration and test program, to verify the requirements before launch, for operations planning, and to assess the on-orbit performance. 

The optical performance requirements were managed through the wavefront error budget that was based on the image quality metrics of Strehl ratio and encircled energy (see Section~\ref{tab:RequirementSpace}). The top-level image quality requirements put direct constraints on the system RMS wavefront error (WFE). Specifically, the driving requirement of a Strehl ratio of 0.8 at 2~$\mu$m wavelength required a system-wide RMS WFE $<$ 150~nm. A portion of this WFE budget was then allocated to each subsystem. This top-level system-level performance was used to make allocations to each of the sub-systems (e.g., telescope, integrating structure, science instruments). The driving encircled energy stability requirement was $<$3\% in a 80~mas radius aperture at 2$\mu$m wavelength over 14 days following a worst case hot-to-cold slew. The encircled energy stability was budgeted by assigning wavefront error allocations in different spatial frequency composition regimes based on capability of the active wavefront control. The low-frequency modes were allocated based on the active global-alignment aberration control, while the mid-frequency modes were allocated based on the active primary mirror figure control. The high-frequency modes were based on the static high spatial frequency errors that were controlled during the fabrication process. Each sub-system's WFE allocation was further divided into WFE static residuals, WFE stability, and image motion. As a system, each allocation could either be tested (e.g, mirror static high-frequency errors) or assessed via integrated modeling (e.g., OTE stability). 

JWST's performance verification could not follow the traditional NASA paradigm to ``test as you fly'' (TAYF). The test facilities were not available for end-to-end testing of an observatory of this size operating at cryogenic temperatures. Rather, the integration and test program developed for JWST relied on incremental performance verification with testing at the sub-system level and, to the extent possible, testing at higher levels of assembly. The test conditions across the observatory were dramatically different. For example, the telescope and science instruments were cryovacuum tested at their operational temperatures of $\sim$40~K in Chamber A at NASA's Johnson Space Center (see Section~\ref{subsubsec:OTIS-cryovac}) whereas the sunshield and spacecraft bus were cryovacuum tested at temperatures ranging between 150--330~K in a more conventional thermal-vacuum environment at Northrop Grumman \citep{McElwain+Niedner+Bowers+etal_2018}. This piecewise verification relied on subsystem test data to anchor the high-fidelity integrated models used to make performance predictions for many of the system-level requirements. 

\subsubsection{Integrated Modeling}

\begin{figure*}
  \centering
%  \vspace{-0.7in}
  %\hspace{-1.5in}
  \includegraphics[width=0.8\textwidth]{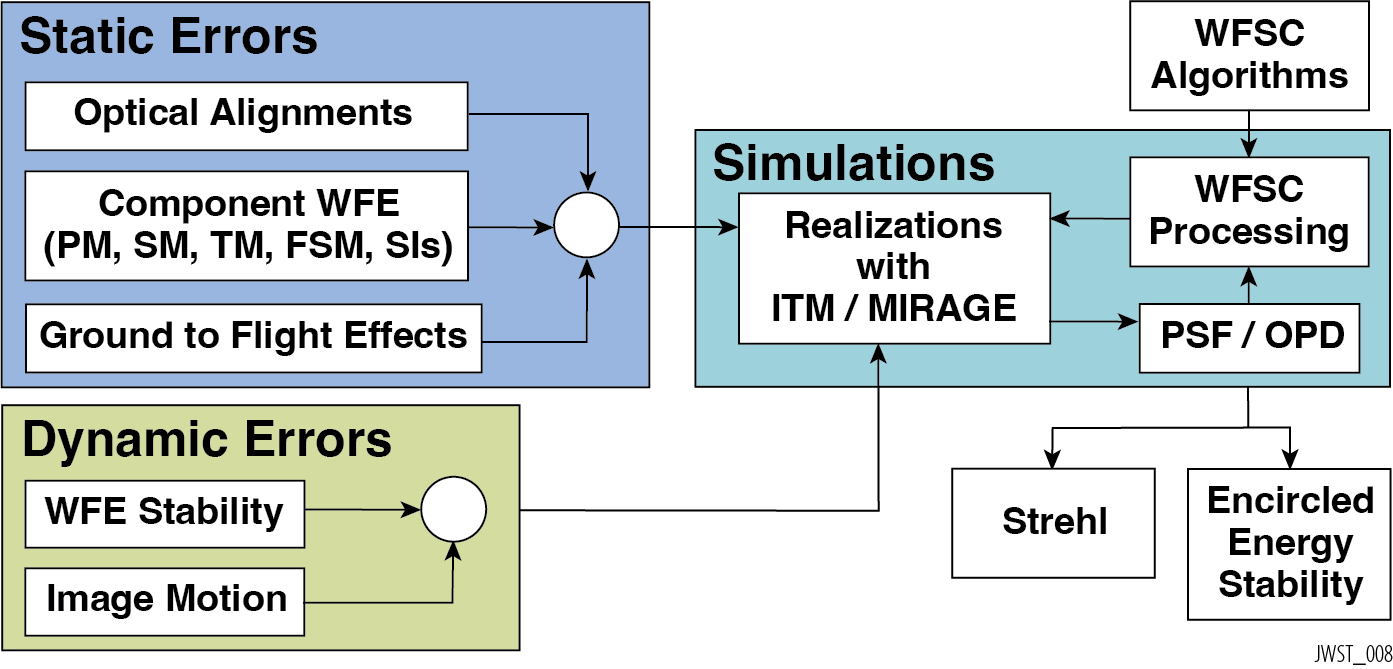}
%   \vspace{-0.9in}
  \caption{JWST's optical performance estimates were made using a combination of test data and integrated models. For requirement verifications, there were conservative model uncertainty factors applied and the end of life performance during worst case conditions were assumed.} 
  \label{fig:IntegratedModeling}
\end{figure*}

Highly complex integrated modeling was a key enabling capacity throughout all stages of development. The optical performance estimates made use of ground test data, integrated models, and simulations including the uncertainties in wavefront sensing and control to verify the pre-launch requirements (Figure~\ref{fig:IntegratedModeling}). Component design and ground test performance was used to provide predictions for optical alignments, component-level wavefront error, and ground-to-flight effects. The dynamic components of the error budget used test data as inputs to an extensive structural-thermal-optical (STOP) integrated modeling process that predicted wavefront stability \citep{Knight+Lightsey+Barto_2012} and line-of-sight image motion \citep{Johnson+Howard+Mosier+etal_2004}. The integrated modeling for telescope performance made use of models of the structure, deployed thermal, thermal distortion, optical performance, dynamics and attitude control, and stray light. Each model was validated upon test data and conservative model uncertainty factors were applied to bound the worst case performance. Image motion predictions made use of exported vibrations and a model of the deployed dynamics of the observatory. The telescope thermal distortion and pointing stability following a worst-case slew used a thermal model that balanced the steady state at the hot attitude (pitched towards the sun) and the cold attitude (pitched away from the sun) with a worst case roll. The small temperature changes, less than 15~mK, from those thermal extremes was predicted for each of the thermal nodes on the telescope and used to determine the mechanical displacements on the structure. The repository of test data and integrated modeling results were used as inputs to the Integrated Telescope Model (ITM) simulator developed by Ball Aerospace in order to predict the optical performance, simulate data products for the development of analysis tools, and to rehearse the telescope alignment process \citep{Knight+Acton+Lightsey+etal_2012}.

The same optical models were also used to inform pre-flight modeling of point spread functions \citep[in particular using the software package WebbPSF,][]{Perrin+Sivaramakrishnan+Lajoie+etal_2014}, which were used extensively in science planning, in development of proposal planning tools such as the exposure time calculator, and in development of data analyses pipelines. The core Fourier optical simulations of PSFs were augmented over time to become part of comprehensive high-fidelity data simulators, such as MIRAGE for NIRCam, NIRISS, and FGS data \citep{Chambers2019AAS...23315712C,Hilbert2022ascl.soft03008H} and MIRISim for MIRI \citep{Klaassen2021MNRAS.500.2813K}. These, along with ITM, became critical enabling tools for the long campaign of preflight rehearsals of the WFSC alignment process (Section~\ref{subsec:wfsc}).

% Integrated Modeling

\subsection{New Technologies Needed}
\label{subsec:NewTechnologies}
% From Technology Non-Advocate Review (T-NAR)

Early in the mission development, three new telescope technologies were identified that needed to be developed explicitly for JWST. The project made significant early investments in these enabling technologies to ensure they would be at a technology readiness level (TRL) of 6 (TRL-6 = fully functional model or prototype, demonstrated in a relevant environment) or higher prior to the mission's preliminary design review (PDR). The telescope technology maturation program included primary mirror segment assemblies, large precision cryogenic structures, and wavefront sensing and control. In addition, new metrology capabilities were needed to verify the performance, and a series of new interferometric techniques were developed to test the stability of the composite structure and verify the optical performance of the telescope at ambient and cryogenic temperatures \citep{Saif+Feinberg+Keski-kuha_2021}. The following discussion provides highlights from the technology development program that was completed in 2006. 

\subsubsection{Primary Mirror Segment Assemblies} Low areal density mirrors were recognized as a key technology gap to enabling a $\sim$25~m$^{2}$ aperture space telescope. The areal density of the Hubble primary mirror is 240 kg m$^{-2}$, while JWST's objective was $<$ 26.5 kg m$^{-2}$, which was achieved. A mirror technology development program was convened to evaluate and advance mirror technologies through multiple programs, including the subscale Beryllium mirror demonstrator (SBMD, \citealt{Reed+Kendrick+Brown+etal_2001}) and the advanced mirror system demonstrator (AMSD, \citealt{Stahl+Feinberg+Texter_2004}). The AMSD program evaluated ULE and Beryllium with a wide variety of parameters such as the optical performances achieved, control authority, mounting, and fabrication schedule. While ULE was deemed to have programmatic advantages, it was found to have an astigmatism as it cooled that was non-deterministic and would have added uncertainty to the development. The decision to select O30 Beryllium, a more isotropic form of Beryllium not previously used in space telescopes, was made following the recommendation from the Mirror Recommendation Board. Beryllium was selected largely due to its small coefficient of thermal expansion (CTE) within the telescope's operational temperature range, making it particularly advantageous during the cryo-polishing fabrication process and achieving the telescope's passive stability objectives by not requiring active thermal control. Beryllium is also light weight, advantageous given the very tight mass constraints for the telescope \citep{Feinberg+Clampin+Keski-kuha+etal_2012}. Beryllium mirrors have flight heritage from previous space missions, including the Spitzer, the \textit{Infrared Astronomical Satellite} (IRAS), and the \textit{Cosmic Background Explorer} (COBE).

\begin{figure}
  \centering
%  \vspace{-0.7in}
  %\hspace{-1.5in}
  \includegraphics[width=0.5\textwidth]{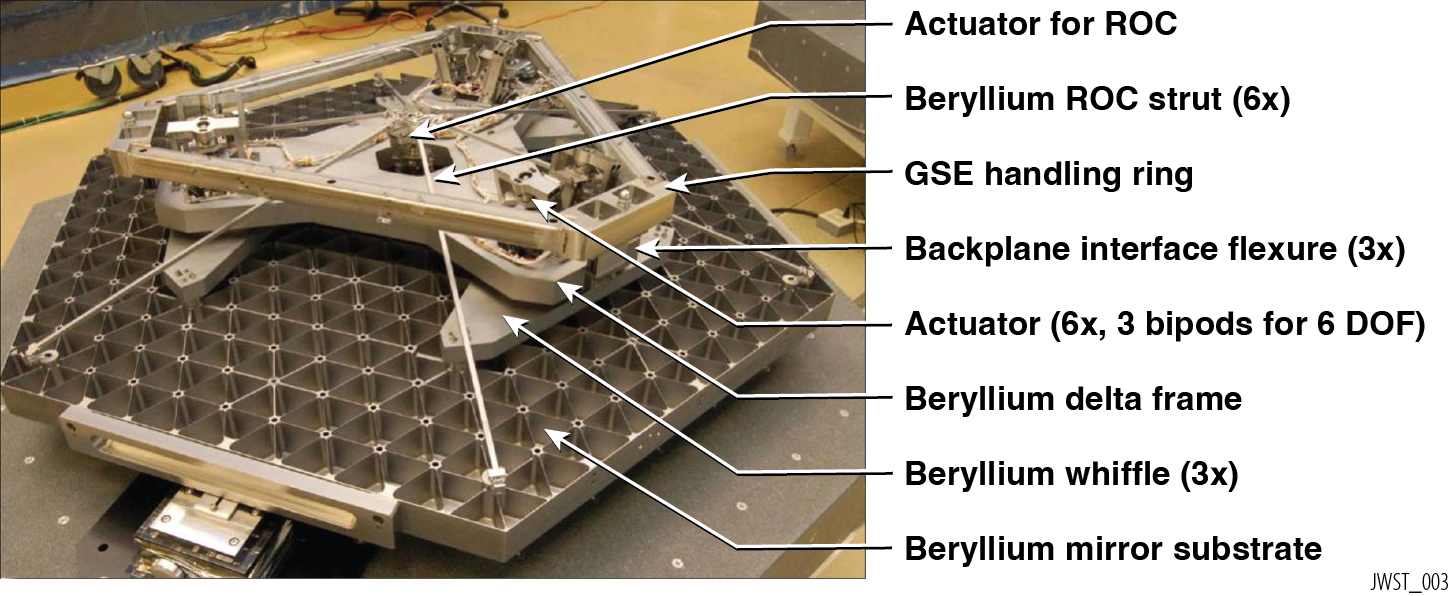}
%   \vspace{-0.9in}
  \caption{Each primary mirror segment and the secondary mirror are supported by a hexapod as shown above. The primary mirror segments also have a radius of curvature actuator that enables the focal lengths for each segment to be matched during the alignment process.} 
  \label{fig:PMSASupport}
\end{figure}

Active control of primary mirror segment position was achieved using actuators mounted in a hexapod arrangement, plus a center actuator for active control of radius of curvature (see Figure~\ref{fig:PMSASupport}). Specialized actuator mechanisms were developed specifically for JWST in order to enable the active positioning of the large segmented mirrors and to support the mirrors during ground test and the launch environment. Each mechanism makes use of a fine stage flexure and coarse drive coupling to control the linear displacement \citep{Warden_2006}. The actuators themselves have remarkable performance parameters including a fine step size of 7.7~nm resolution, with 2~nm of fine repeatability, over a 10~$\mu$m fine range. A coarse drive coupling in the same mechanism provides a 58~nm step size over a full 21~mm. Further, unlike ground telescope active and adaptive optics, which often use electrostatic or piezoelectric actuators, JWST's actuators operate mechanically via a gear train and flexures; the mechanical gear trains hold position stiffly and precisely even when the actuator is entirely unpowered, which is necessary to avoid undesired waste heat into the cryogenic telescope. 

% Hexapod

\begin{figure}
  \centering
%  \vspace{-0.7in}
  %\hspace{-1.5in}
  \includegraphics[width=0.4\textwidth]{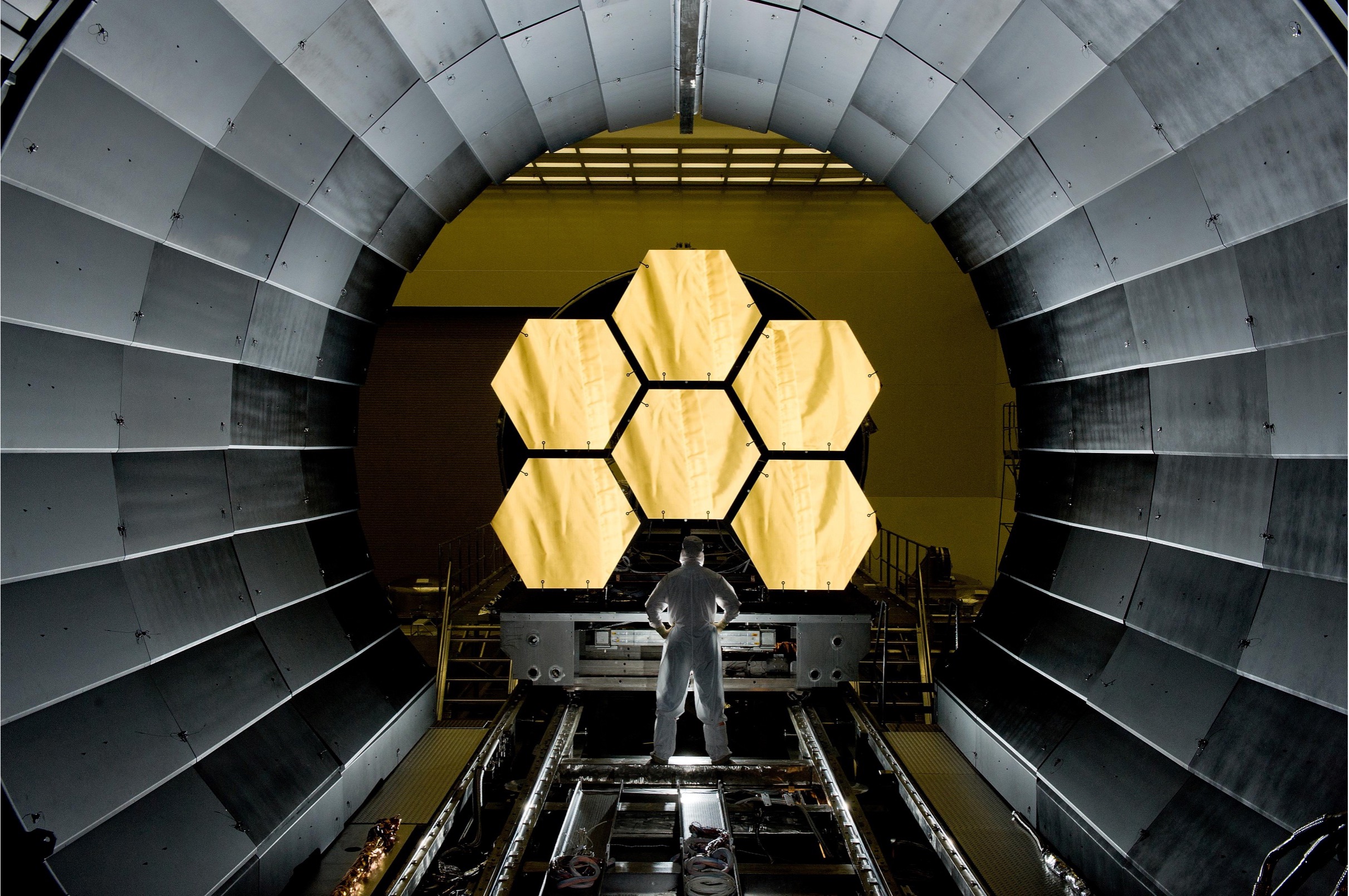}
%   \vspace{-0.9in}
  \caption{Six of the flight mirror segments prepared to undergo acceptance testing in the XRCF. The segment-level optical performance and structural stability was precisely measured.} 
  \label{fig:XRCF-Testing}
\end{figure}

The primary mirror segment development process required new facilitization and made use of economies of scale to fabricate the multiple segments in parallel. The mirror blanks were made from O30 Beryllium through a hot isostatic pressing process by Brush Wellman. The blanks were then light weighted by removing over 92\% of the material in a honeycomb shape by Axsys Technologies. Next, the mirror substrates were polished by Tinsley Labs and each mounted to its flexure and radius of curvature system. The mirrors were then optically tested at ambient and cryogenic temperatures at X-ray and Cryogenic Facility (XRCF) at NASA's Marshall Space Flight Center, followed by another round of cryo-polishing to ensure each mirror achieved the correct optical figure at the intended cryogenic operating temperature \citep{Cole+Garfield+Peters+etal_2006}. The mirrors were gold coated using a vacuum vapor deposition process by QCI, Inc. The gold coating provides high reflectivity across the operational wavelength range of 0.6--28.1$\mu$m \citep{Keski-kuha+Bowers+Quijada+etal_2012}. A protective SiO$_x$ overcoat was applied that improved the durability of the coating and enabled cleaning at stages throughout the I\&T process (see \citealt{Lobmeyer+Carey_2018}). Finally, flight acceptance testing for each segment was carried out in the XRCF facility (see Figure~\ref{fig:XRCF-Testing}). 

% XRCF Testing

\subsubsection{Large Precision Cryogenic Structure} 
A large precision cryogenic structure was necessary to enable the passive optical stability without active control. The telescope and the science instruments are supported by a composite optomechanical structure that must withstand the launch environment loads, deploy within the capture range of the mirror actuators, survive the stresses induced from cooling down to cryogenic temperatures, and have minimal thermal distortion. 

\begin{figure}
  \centering
%  \vspace{-0.7in}
  %\hspace{-1.5in}
  \includegraphics[width=0.3\textwidth]{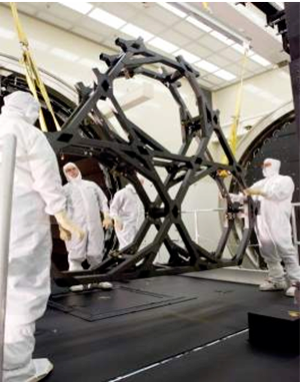}
%   \vspace{-0.9in}
  \caption{The Backplane Stability Test Article (BSTA) was an engineering model for 1/6 of the JWST backplane, including the structural elements used to create the entire backplane assembly. This shows the BSTA being prepared for cryotesting at operational temperatures in the XRCF.} 
  \label{fig:BSTA}
\end{figure}

Early in the JWST development, it was recognized that the materials database for composite structures was inadequate over the operational temperature range, the ability to measure deformations was inadequate for the JWST verification testing, the engineering modeling tools needed development, and manufacturing process controls needed improvement. The coefficient of thermal expansion for the materials needed to be measured to less than 30 ppb K$^{-1}$ at temperatures of $<$ 25 K, which was more than 100 times more precise than the state of the art at the time \citep{Atkinson+Arenberg+Gilman+etal_2007}. ATK implemented a technique to measure the CTE for large structures with the precision needed for JWST. With the materials characterized, a prescription for the composite structure was defined that used unidirectional prepreg made from M55J carbon fibers and resins into laminant mixtures tuned for the appropriate strength and thermal performance. Manufacturing controls were established to precisely align the fibers during layup and closely manage the fiber to resin ratio necessary for precise material properties. Controls were also put in place to achieve the desired bonded joint adhesive thickness used to connect the individual tubes into a truss. A prototype of the composite structure, called the Backplane Stability Test Article (BSTA), was built by ATK and tested at MSFC's XRCF (Figure~\ref{fig:BSTA}). Verification of the structure's stability made use of a new Electronic Speckle Pattern Interferometer (ESPI) metrology approach \citep{Saif+Bluth+Greenfield+etal_2008}, a technology development in itself, to confirm the structure was TRL-6.

% BSTA

\subsubsection{Wavefront Sensing \& Control} 

\label{subsec:wfsc}
%7~nm resolution with $\sim$12.5~mm range

The image-based phase retrieval methods used with JWST have a heritage stretching back to the diagnosis and correction of the infamous spherical aberration in the Hubble primary mirror \citep{Krist_1995ApOpt..34.4951K,Fienup_1993ApOpt..32.1747F}. Phase retrieval using the science instruments elegantly avoids the need for substantial dedicated wavefront sensing hardware, and ensures sensing of wavefronts directly at the science focal planes. However, to accommodate the evolving alignment of the mirrors (from initial deployment errors measured in millimeters to final alignments measured in nanometers) these methods must operate over a tremendous dynamic range, and must also sense dissimilar and degenerate degrees of freedom. As a result several distinct wavefront sensing methods must be used. The primary tool is focus-diverse phase retrieval, using a hybrid diversity algorithm developed specifically for JWST \citep{Dean_2006SPIE.6265E..11D}. Focus diversity is provided at different stages by defocusing the secondary mirror or by using weak lenses within NIRCam that can be inserted into the beam path. The focus diversity method is augmented with dispersed Hartmann sensing for the measurement of segment piston \citep{Shi+Chanan+Ohara+etal_2004}.

\begin{figure}
  \centering
%  \vspace{-0.7in}
  %\hspace{-1.5in}
  \includegraphics[width=0.4\textwidth]{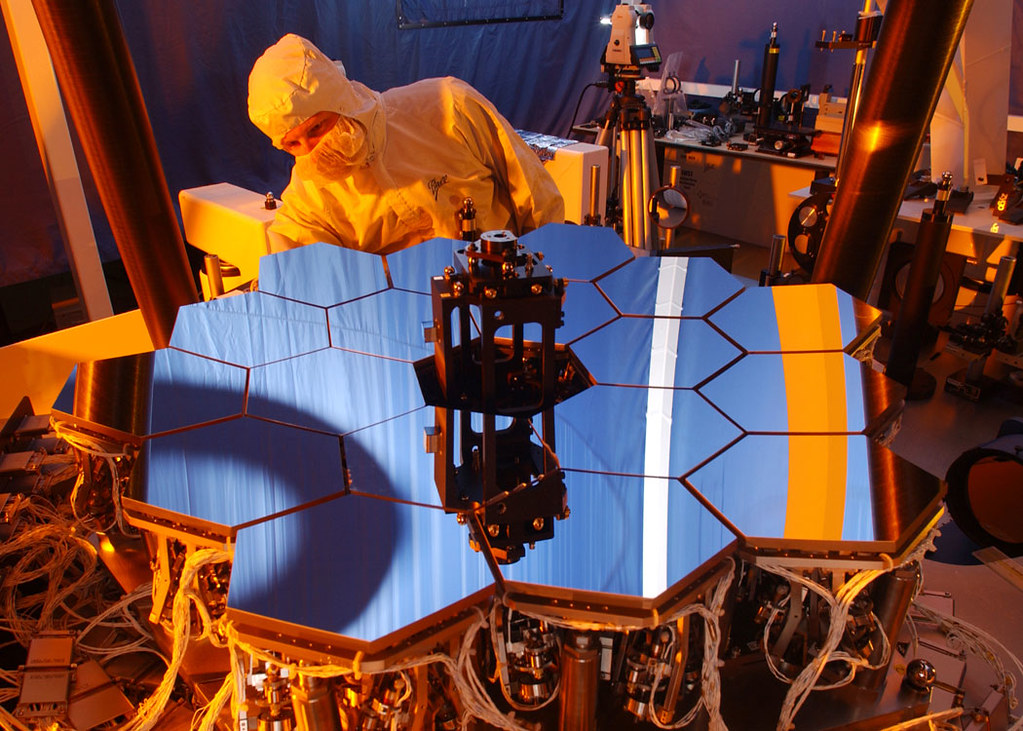}
%   \vspace{-0.9in}
  \caption{The JWST Test Bed Telescope (TBT) is a 1:6 scale model of JWST's telescope with the full sensing capabilities and control authority, located at Ball Aerospace.} 
  \label{fig:TBT}
\end{figure}

The step-by-step sequence of sensing and control activities, as well as the associated algorithms and software, were developed at Ball Aerospace. To test and prove the implementation, a 1:6 scale model and functionally accurate Test Bed Telescope (TBT) was built (\cite{Kingsbury_2004SPIE.5487..875K}, see Figure~\ref{fig:TBT}). Using the TBT, the complete end-to-end telescope alignment process was successfully demonstrated, achieving TRL~6 in 2006 \citep{Acton_2007SPIE.6687E..06A,Feinberg_2007SPIE.6687E..08F}.

% Testbed Telescope

Even with those fundamental tools proven, a decade of work remained to mature them from lab-scale demonstration to flight-ready processes. Operational implementation of the commissioning plan was complicated by the need to begin operation of fine guidance control while still adjusting mirrors (see Section~\ref{subsection:Guiding&LOSPointing}), and to interweave telescope alignment with prerequisite steps of instrument commissioning such as focal plane calibrations (see Section~\ref{subsection:FPCals}). The methods were refined and operational plans were prepared leading up to launch \citep{Perrin+Acton+Lajoie+etal_2016}, culminating in detailed implementation plans, procedures, and observing programs. Contingency plans were prepared in the event that nominal plans could not be followed, for many distinct contingency scenarios. Phase retrieval analysis software was similarly iteratively refined prior to launch; the initial delivery of flight WSS software occurred in 2011, and regular improvements continued thereafter as part of the I$\&$T of the Science $\&$ and Operations Center (SOC). 

These processes for WFSC were repeatedly tested together with the flight hardware at various stages of observatory I\&T. The initial test and operation of the integrated OTE electronics and mechanisms took place in 2016 (see Section~\ref{sec:otis_ambient}). During the cryovacuum test of the telescope plus instrument suite (Section~\ref{subsubsec:OTIS-cryovac}) while the majority of that test program used GSE for metrology of observatory alignments, specific activities were included to test the flight scripts for wavefront sensing and control using flight hardware. This was the first and only time that NIRCam was used to sense OTE mirror alignments on the ground \citep{Acton+Knight+Chonis+etal_2018}. In parallel, the data generated by that activity was flowed back to STScI and used for a demonstration of sensing and control software processes using the integrated SOC. This was the first major demonstration of processing of JWST data in a flight-like manner at the mission operations center \citep{Lajoie+Perrin+Myers+etal_2018}. The commanding for mirror moves and deployments was repeatedly exercised as part of regular OTE functional checks, up to and including at the launch site. 

A necessary input to the WFSC process was accurate knowledge of instrument-specific wavefront errors, to allow subtracting the instrument contributions from the results of the image-based phase retrieval to perform OTE wavefront control. This objective was met through precise measurements at dozens of field points within all instruments, as part of instrument cryo-vacuum testing completed by 2016. 

The wavefront sensing and control activities demanded human-in-the-loop controls and required training a large wavefront team for round the clock operations during commissioning. The telescope was aligned start-to-finish over a hundred times in simulation, individually by many members of the team and in collaboration. As part of this training process, there were iterative refinements of the methods, procedures, and documentation. The individual simulations built towards larger team rehearsals, including 20 internal wavefront team practices and 25 mission operations or science operations team wide rehearsals. Many of these rehearsals were carried out throughout the COVID-19 pandemic under work-from-home conditions, remotely. The extensive rehearsal program was a critical, invaluable activity in building a smooth-functioning cohesive wavefront team combining staff from multiple organizations and skillsets.

\subsection{Integration \& Testing} \label{subsec:IntTest}

The JWST integration and test program began at the component level and continued as the hardware was integrated into successively larger sub-assemblies, assemblies, and eventually the entire observatory system. At various levels of assembly, testing took place to confirm the functional performance and verify the workmanship. The testing approach followed a philosophy of independent, incremental testing with predefined success criteria, crosschecks that included end-to-end testing, comprehensive external and internal review, thorough risk management, and open transparent communications and documentation. A strict requirement of the optical test program was that the optical test equipment in a verification test had to be independent of the test equipment used to fabricate the optics \citep{Feinberg+Geithner_2008}.

\begin{figure*}
  \centering
%  \vspace{-0.7in}
  %\hspace{-1.5in}
  \includegraphics[width=0.9\textwidth]{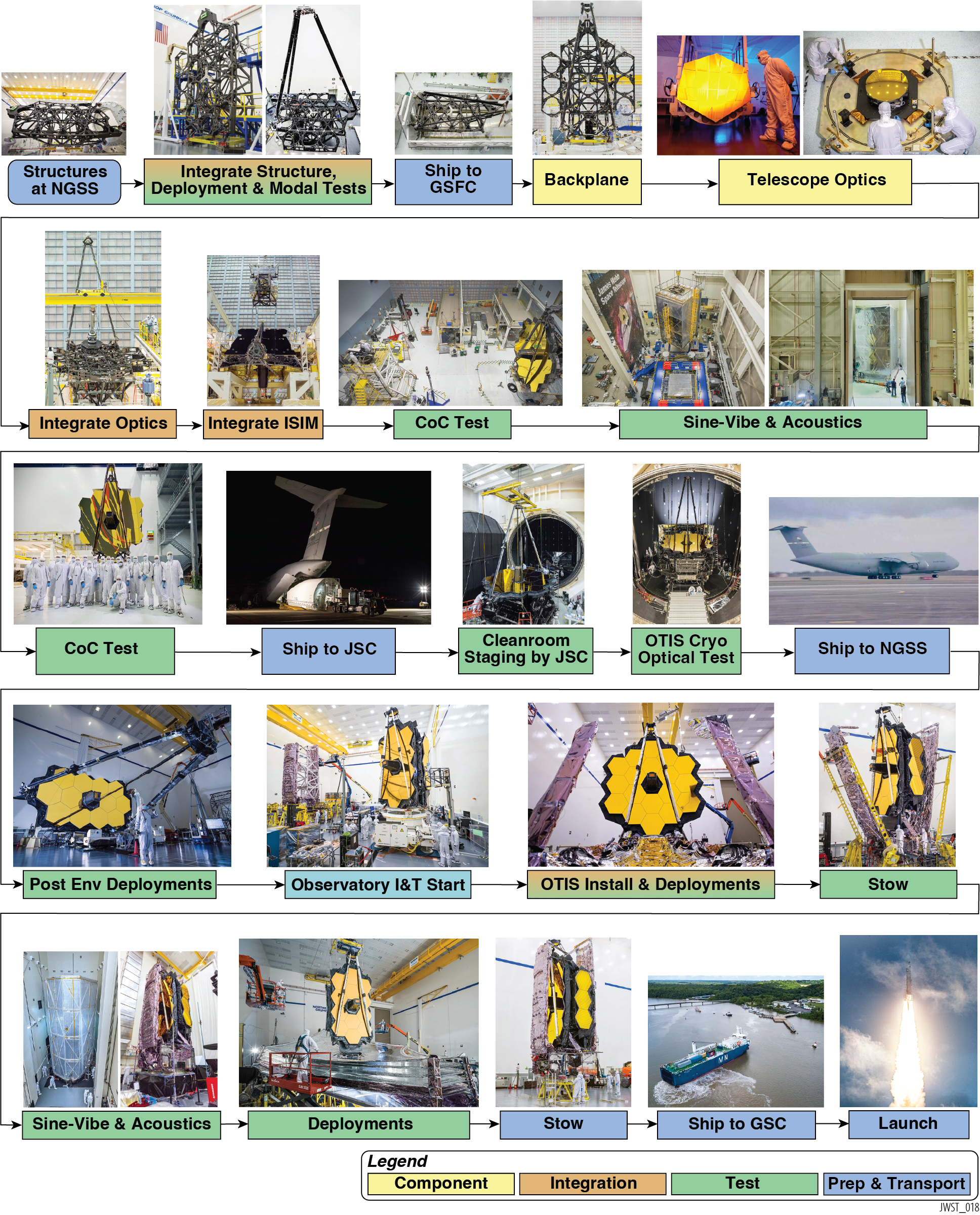}
%   \vspace{-0.9in}
  \caption{A high level overview of the telescope integration and test sequence through launch.} 
  \label{fig:OTIS_Flow}
\end{figure*}

In this section we highlight some key activities from the later stages of the I\&T program for the OTE. We then discuss the major I\&T activities for the OTIS, which is the term for the combination of the OTE plus the Integrated Science Instrument Module (ISIM); OTIS = OTE + ISIM. We conclude the section with a summary of the activities involving the OTIS after delivery back to Northrop for integration with the rest of the observatory. A high-level flow of the OTE activities through its integration and testing through Observatory readiness for launch is presented in Figure~\ref{fig:OTIS_Flow}.

% OTIS Flow 

\subsubsection{Flight OTIS Integration \& Ambient Testing}

\label{sec:otis_ambient}

The major structural elements of the OTE (see Figure~\ref{fig:TelescopeExplodedView}), as well as the associated electrical harnesses, were integrated and tested at Northrop in 2014--2015. Major tests included load testing of the mirror backplane and testing of the Deployable Tower Assembly (DTA). After precision integration of the DTA, the two telescope wing structures, and the Secondary Mirror Support Structure (SMSS) to the backplane, these subsystems were exercised for functionality and repeatability with ambient deployments \citep{Glassman+Levi+Liepmann+etal_2016}. Modal surveys were also carried out in the stowed and deployed configurations, with appropriate mass simulators for hardware to be integrated later.

\begin{figure}
  \centering
  \includegraphics[width=0.4\textwidth]{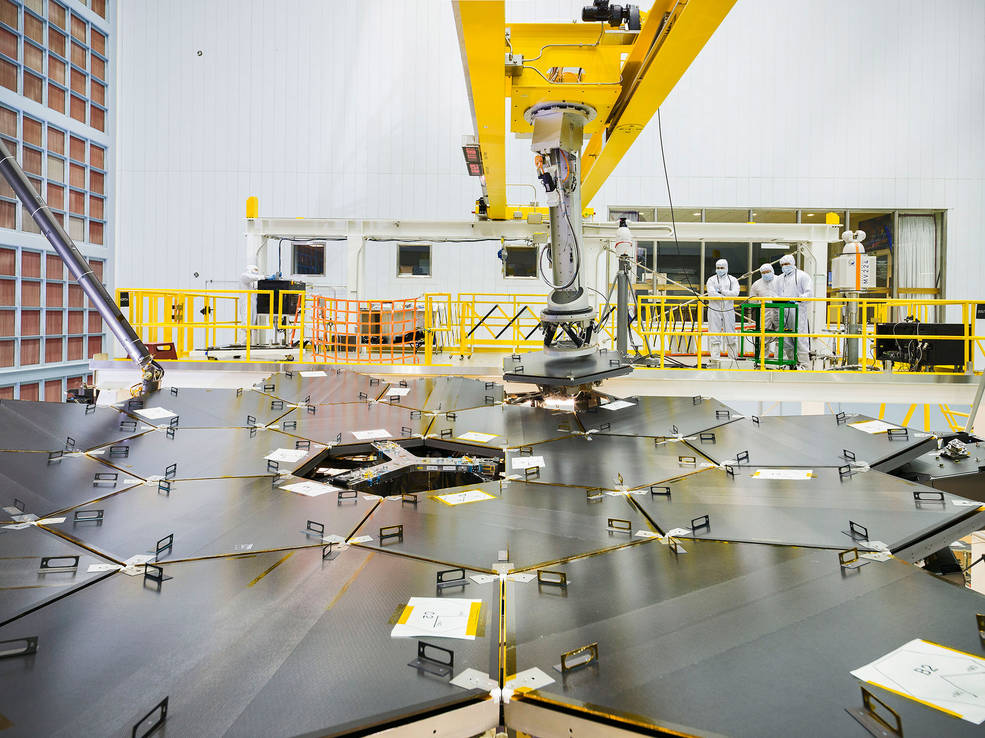}
  \caption{Installation of the final Primary Mirror segment, assisted by a high-precision robotic arm, suspended from traveling stages. Black covers were mounted on the mirrors to protect from contamination.} 
  \label{fig:PM installation}
\end{figure}

After shipping of this hardware to GSFC, the integration of additional harnesses and small electronics boxes (e.g. for the mirror actuators) and the OTE optics took place in a dedicated assembly and alignment facility in GSFC’s largest cleanroom. The PMSA shim prescription was determined using metrology from a coordinate measuring machine brought from Tinsley Labs, now Coherent Inc., and laser tracker measurements of the composite backplane structure. The PM segments were mounted to the backplane with the assistance of a traveling robot arm, as shown in Figure~\ref{fig:PM installation}. Laser trackers measured the alignment state to guide the installation, with laser radar independently measuring. Custom ground shims and adhesive-filled pin gaps secured the location to mechanical tolerances that were a small fraction of the range budget for the PM actuators (see \citealt{Atkinson+Texter+Keski-kuha+etal_2016} for details).

% OTIS Primary Mirror Installation

Integration of the Secondary Mirror Assembly and the Aft Optics Subsystem (AOS) completed the OTE. With the integration of the Fixed and Aft Deployable ISIM Radiators and the ISIM itself (the instrument module and the associated electronics compartment), and an array of thermal and stray light control blankets ($>$900!), the OTIS was complete. The ISIM of course had been through its own comprehensive I\&T program, including three cryo-vacuum tests totaling nearly 300 days of round-the-clock operations \citep{Kimble+Vila_VanCampen+etal_2016}. This integration of the OTE + ISIM to form the OTIS took place in May 2016. 

Using the fully assembled OTIS, the integrated mirror control system hardware and software was exercised, first in small steps starting in October 2016, and eventually in partial deployments of all 18 PMSAs and the SM in preparation for ambient optical testing. This included operation of PMSA actuators and sensors (resolvers and linear variable differential transducer [LVDT] length sensors), operated by the Actuator Drive Unit electronics and controlled by the Wavefront Sensing System software. One minor anomaly discovered during this time is that a small number of LVDT sensors do not operate nominally. This was accepted to ``use as is'', given the availability of other telemetry to confirm mirror motions for those segments (e.g. using resolver telemetry). For some of the affected LVDT sensors, a modified operations concept was developed that used temperature-dependent calibrations to make the sensor information usable, later used successfully in flight. Mirror control mechanisms and processes continued to be exercised throughout the remainder of OTIS I\&T, in particular during the OTIS cryovacuum test.

% OTIS Ambient CoC Test
\begin{figure}
  \centering
  \includegraphics[width=0.4\textwidth]{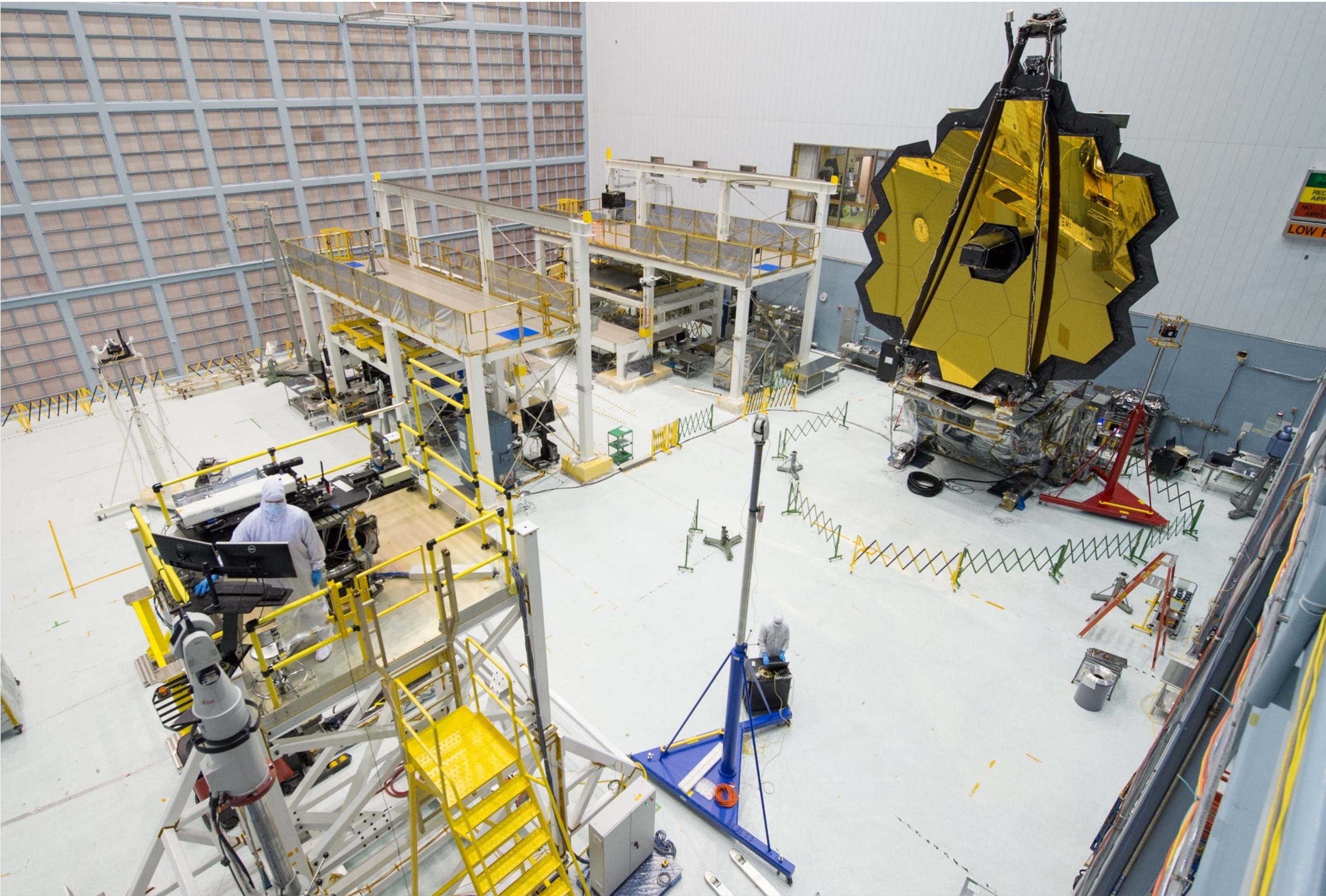}
  \caption{Center of Curvature test setup for ambient optical measurements of the PM. The high-speed interferometer, null lens, and CGH are mounted on the stable platform at lower left.} 
  \label{fig:Center of Curvature setup}
\end{figure}

The OTIS underwent proto-flight level ambient environmental testing (vibration and acoustics) at GSFC in 2016--2017. Ambient optical measurements were carried out before and after those mechanical tests utilizing a Center of Curvature test setup, including a null lens and Computer Generated Hologram (CGH, for working with the aspheric PM surface), as illustrated in Figure~\ref{fig:Center of Curvature setup}; see \cite{Saif+Chaney+Greenfield+etal_2017}. The Center of Curvature setup incorporated a high-speed interferometer for making figure measurements at rates up to 5.9~kHz. With this equipment, the static wavefront of the PM segments was measured before and after the mechanical environmental testing, along with the dynamic response of the backplane and mirror mounts. With a vibration stinger to excite the payload, mechanical transfer functions were measured to look for any signs of damage after the vibe and acoustics tests – e.g. cracks in the structure; loosening of joints. No such damage was seen, with the figure and dynamic measurements repeating pre- and post-test within expected tolerances. Electrical functional checks were also carried out before and after the mechanical environmental testing, along with ``first motion'' (flinch) tests of deployment systems that couldn't be fully deployed at GSFC in the one-g environment.

An anomaly did arise during the OTIS-level vibration test, when a loud ``bang'' was heard. This was determined to have been caused by gapping at one of the preloaded interfaces of a Launch Release Mechanism in a PM mirror wing. Modifications to the procedures to properly set and maintain the preload of these interfaces resolved this issue, and no damage had been done. In addition, excessive resonant response was seen for the SMSS and the AOS at some frequencies due to the low damping of the large composite structure. The test vibration spectrum was notched at these frequencies to protect the hardware during the OTIS-level test. Particle dampers were subsequently designed and installed onto the AOS and SMSS to reduce these responses; they operated successfully later at Observatory-level testing and ultimately through the actual launch.

% OTIS cryovacuum test
\begin{figure*}
  \centering
  \includegraphics[width=0.75\textwidth]{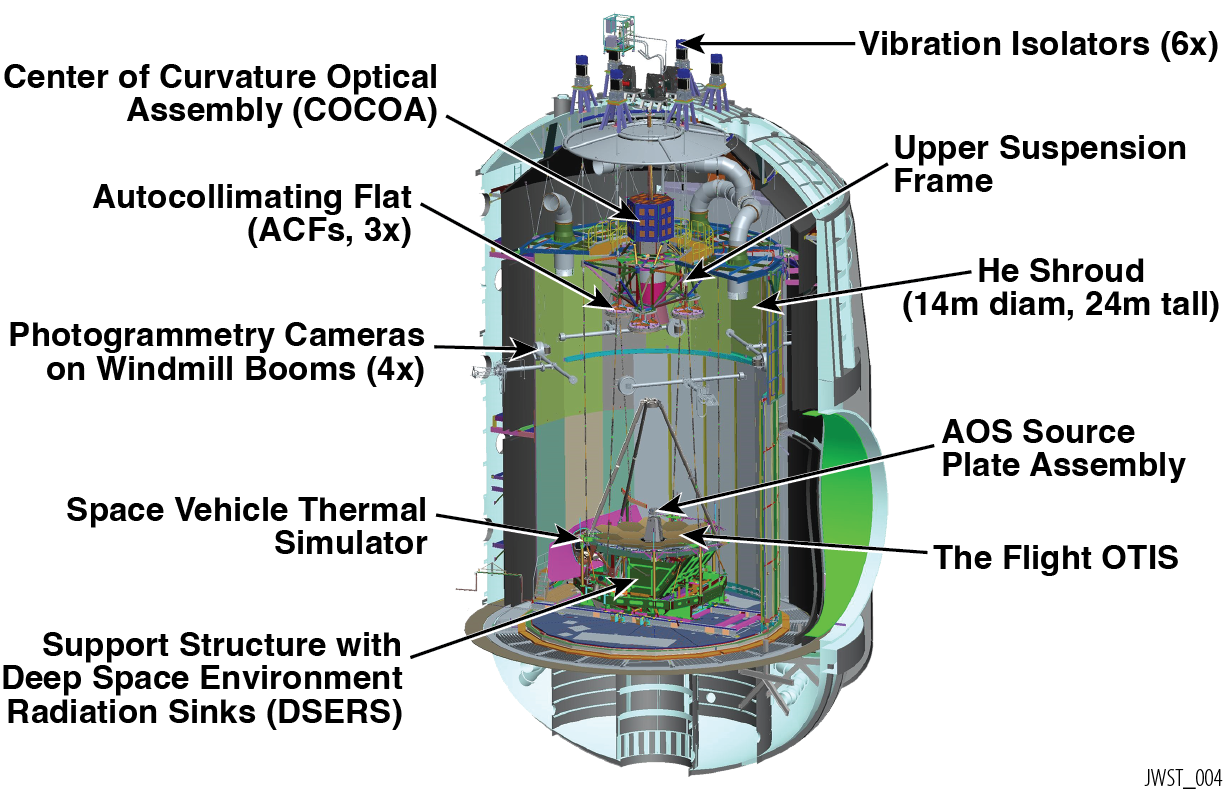}
  \caption{OTIS cryovacuum test configuration showing the telescope inside the chamber with optical metrology test equipment.} 
  \label{fig:OTIS test config}
\end{figure*}

\subsubsection{Flight OTIS Cryo-Vacuum Test} \label{subsubsec:OTIS-cryovac}

The final phase of I\&T for the OTIS was an extraordinarily challenging cryo-vacuum test, previously described in \cite{Feinberg+Barto+Waldman+etal_2011} and \cite{Kimble+Feinberg+Voyton+etal_2018}. This took place at NASA's Johnson Space Center in historic Chamber A, which is a US national historic landmark from the Apollo program. After shipment to JSC, the OTIS underwent electrical functional testing, the SMSS was deployed (with assistance -- the necessary GSE for a powered deployment in one-g was only at Northrop), and the DTA and PM wings were deployed. 

After these activities, the payload was configured for the cryo-vacuum test in Chamber A, which had been extensively refurbished for the thermal and contamination requirements of JWST and outfitted with optical and thermal GSE comprising an elegant test architecture, illustrated in Figure~\ref{fig:OTIS test config}. This architecture supported a rich array of operational, thermal, and optical test goals, with 40 separate tests. End-to-end optical tests were carried out using sub-apertures of the primary mirror. The test campaign applied lessons learned from the Hubble program.

Key components of the optical test equipment included the Center of Curvature Optical Assembly (COCOA, \citealt{Wells+Olczak+Merle+etal_2010}), whose interferometers had a view of the entire PM; photogrammetry (PG) cameras on four rotating booms \citep{Lunt+Wells+Rhodes+etal_2020}, which provided remarkably accurate relative positions (sub-100 $\mu$m) of optical targets over the many-meter distances involved, through image triangulation; the AOS Source Plate Assembly (ASPA), which mounted light sources (fiber-fed or local) at the intermediate Cassegrain focus of the OTE – these provided downward (half-pass) images through the TM, FSM, and SIs and upward (pass-and-a-half images) through the SM, PM and then, after reflection off sub-aperture Auto-Collimating Flats (ACFs), back through the entire OTIS optical train; and fiducial light strips straddling the edges of the PM. The position of the ASPA sources made their images highly aberrated, but in a precisely known way, so that alignments and OTIS wavefront measurements could be extracted nonetheless. The downward sources were used to test the guiding control loop.

% OTIS Cryotest Schematic
\begin{figure}
  \centering
  \includegraphics[width=0.4\textwidth]{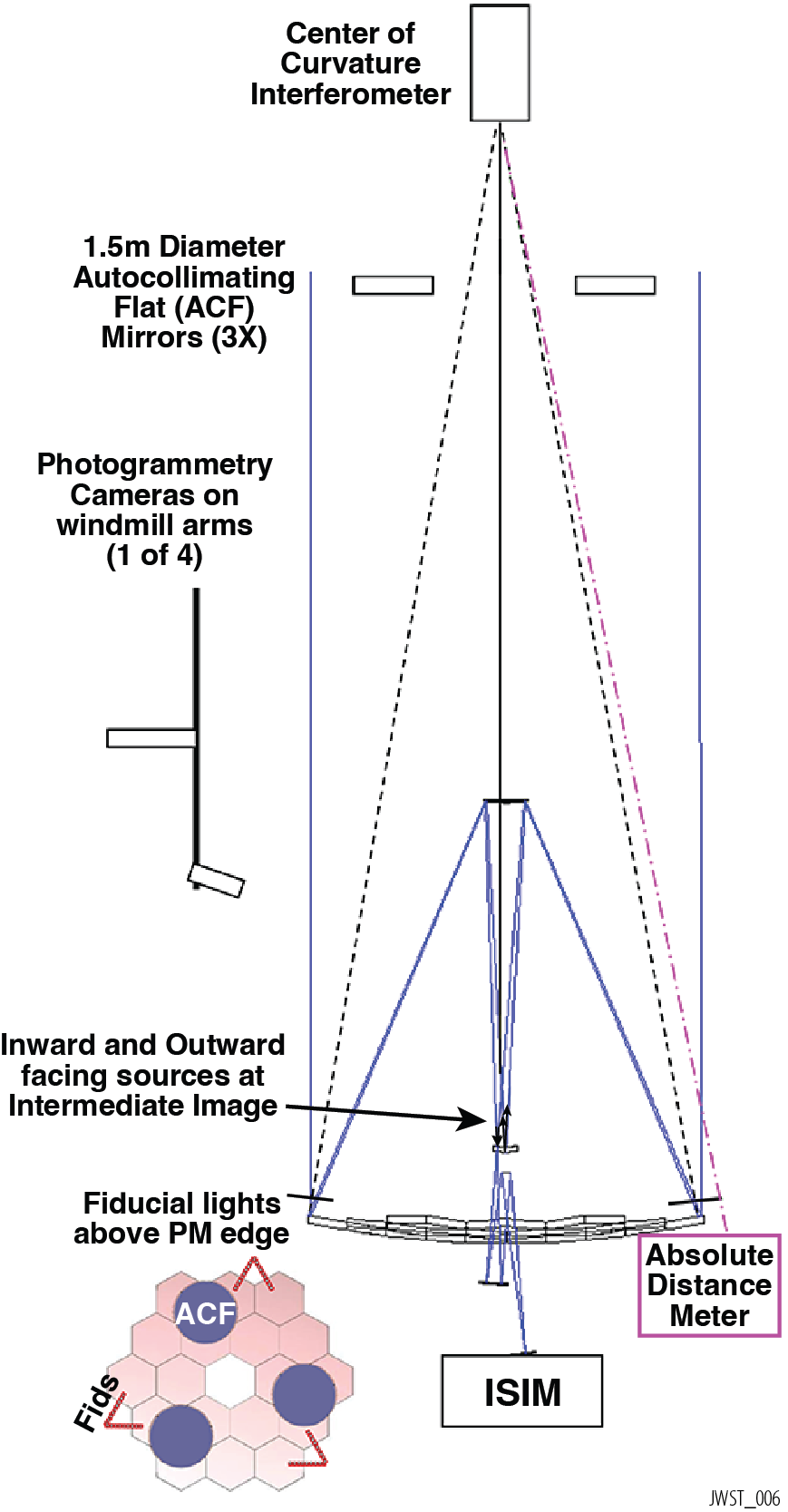}
  \caption{Simplified OTIS optical test schematic.} 
  \label{fig:OTIS optical test schematic}
\end{figure}

The optical test equipment (see Figure~\ref{fig:OTIS optical test schematic}) worked together to satisfy the critical optical verification goals (such as the verification of the non-adjustable AOS to ISIM alignment, verification of the radius of curvature) as well as various cross-check goals. A succinct description of the process is as follows:

\begin{itemize}
\setlength{\itemsep}{0pt}
  \item PM segments were aligned and phased via photogrammetry and COCOA interferometry.
  \item The SM was aligned via photogrammetry and checked with Pass-and-a-Half imaging to the NIRCam instrument.
  \item AOS to ISIM alignment was verified via Half-Pass imaging using inward facing sources and all science instruments.
 \item Fiducial lights above the primary mirror were used for verifying pupil alignment, using NIRCam’s pupil imaging capability.
  \item End-to-end imaging and field tilt was cross-checked using Pass-and-a-Half imaging using outward facing sources, the autocollimating flats, and all of the science instruments.
  \item Wavefront Sensing \& Control hardware checks and demonstrations were performed via Pass-and-a-Half testing and NIRCam.
\end{itemize}

\subsubsection{The Pathfinder Program} \label{subsubsec:Pathfinder}

The Pathfinder Program used spare and test equipment as surrogates for the flight hardware in order to prove out many of the integration and testing activities (see \citealt{Feinberg+Keski-kuha+Atkinson+etal_2014}, Section~\ref{subsec:LessonsHubbleChandra}). The test article, referred to as the ``Pathfinder'', was comprised of two spare primary mirror segments, a spare secondary mirror, and composite structure representative of the center section and the secondary mirror supports. The Pathfinder was transported and integrated using the protocols and procedures that would later be used on the flight hardware, in some cases demonstrating the actual capabilities of the test equipment.

Following integration, the Pathfinder was used to prepare the cryovacuum testing facility and equipment at JSC in 2015/2016, while the OTE and OTIS were being integrated and tested at Goddard. The Pathfinder cryotest program utilized the thermal and optical GSE developed for the OTIS test and the Pathfinder structure itself \citep{Matthews+Kennard+Broccolo+etal_2015}. The first Optical GSE test (OGSE1) checked out the COCOA and PG operations with those systems, while the second (OGSE2), incorporated the \textit{flight} AOS (hence requiring careful coordination with the flight I\&T flow) and the ASPA to dry-run the half-pass and pass-and-a-half tests as well. A third Pathfinder test included thermal mock-ups of the remaining center-section PM segments and validated the cooldown and warmup procedures that would be needed, including the requisite contamination control \citep{Matthews+Scorse+Spina+etal_2015}.

\subsubsection{The Flight OTIS Cryo-Vacuum Test Results} \label{subsubsec:OTIS-cryovac}

Cryo-vacuum testing of the flight OTIS took place in 2017. The 100-day, round-the-clock test campaign was remarkably successful, despite the many challenges, including the passage of Hurricane Harvey through the JSC area in the middle of the test, which devastated the local community and shut down JSC for normal operations with 55" of rainfall. After a 5 day period of special hurricane operations, and nearly running out of liquid nitrogen, the cryotest narrowly missed being abruptly ended. Fortunately the hurricane did not preclude safely continuing the test. All significant goals for the test were achieved, including confirming the health of the OTIS payload after its environmental test program \citep{Wolf+Gallagher+Knight+etal_2018}, accomplishing the planned optical verifications and cross-checks (e.g., \citealt{Hadaway+Wells+Olczak+etal_2018}), validating the OTIS thermal model and the OTIS thermal distortion model, both required for the Integrated Modeling of the observatory as a whole, and various operational validations and demonstrations.

Performance of the OTIS was overall excellent, with predictions that satisfy the mission-level requirements \citep{Lightsey+Knight+Barto+etal_2018}. But, there were several types of optical instabilities identified.  One, caused by over-tightness of the soft-structure frill stray light blocker and PMSA closeouts at the cryogenic operating temperatures (such that they exerted temperature-dependent forces on the mirror backplane), was mitigated with post-test modifications to restore the intended slack where feasible. A second, which coupled temperature variations in heater-controlled radiator panels on the instrument electronics compartment (IEC) to the mirror backplane, inducing oscillating structural distortions, was demonstrated with ambient measurements and analysis to be caused by a rigid \textit{non-flight} mounting of the IEC for the OTIS CV test. Both of these have been shown to have minimal wavefront impact in flight (see Section~\ref{subsubsec:ThermalStability}) 

A final instability, called ``tilt events'' referred to sudden, stochastic changes in the piston/tip/tilt pose of individual PM segments. Several such events were seen throughout the OTIS test period. Though not fully explained, these were ascribed to stick/slip release of stresses from cooldown thermal deformation in the OTIS structure. These reduced during the end of the test, and it was expected that these would fade away with time after cooldown as the various stresses in the system were gradually relieved. This behavior during the OTIS test informed expectations that such events could be seen in flight. This supposition appears to be confirmed by the flight behavior (see Section~\ref{subsubsec:TiltEvents}), though with the excellent sensitivity of the flight wavefront sensing, most of the observed tilt events in flight are actually below the detection threshold of the OTIS cryo-vac analyses.

\subsubsection{OTIS during Observatory I\&T}

After the deconfiguration of the OTIS from the test configuration, it was shipped to prime contractor Northrop Grumman’s Space Park in Redondo Beach, California. While at Northrop, powered deployment of the SMSS and DTA took place with appropriate one-g off-loading hardware. Both deployments were made from the spacecraft electronics to demonstrate the connections and scripts were working properly. For the SMSS, this represented the only post-OTIS-vibration powered deployment, confirming the health of the deployment system after that proto-flight-level vibration exposure.

In the summer of 2019, the OTIS was integrated with the spacecraft and sunshield to form the full-up JWST observatory. Alignment metrology was performed in the integrated configuration to characterize the OTE to star tracker boresights.

In the full-up observatory configuration, the payload underwent various deployment tests (e.g., off-loaded deployment of the DTA and PM wings). The observatory as a whole was then put through acceptance-level vibration and acoustic tests, with subsequent deployment and electrical functional tests. Like the OTIS, the spacecraft and sunshield had previously successfully undergone mechanical environmental testing at proto-flight levels.

Both the PM and the SM were cleaned of particulates at appropriate times in the Northrop flow, with a gentle brush technique described by \citealt{Lobmeyer+Carey_2018}. For the SM, which had the most challenging particulate contamination budget (so cleaning was desired as late as possible), this cleaning took place after the final stowing of the observatory into its transport (and launch) configuration, just before encapsulation of the observatory into a clean, environmentally-controlled shipping container for transport by sea to the Guiana Space Centre (GSC) in Kourou, French Guiana. There, the observatory executed final ground functional tests, was fueled, and was encapsulated in the Ariane 5 rocket fairing for launch.

\section{On-Orbit Commissioning and Characterization}
\label{sec:OnOrbitCharacterization}

Following JWST's launch, the telescope was deployed, aligned, characterized, and readied for science observations.  
In this section, we describe the overall sequence of activities executed during OTE commissioning as well as the resulting performance of the optics, pointing, and focal plane alignment. Many years of preparation, for both the observatory hardware and the commissioning operations plans and teamwork, resulted in a smooth and efficient commissioning which completed successfully and as scheduled, and delivered an OTE performing at or above requirements.

\subsection{OTE Commissioning Activities} \label{subsec:OTECommissioningActivities}

\begin{figure*}
  \centering
  \includegraphics[width=0.8\textwidth]{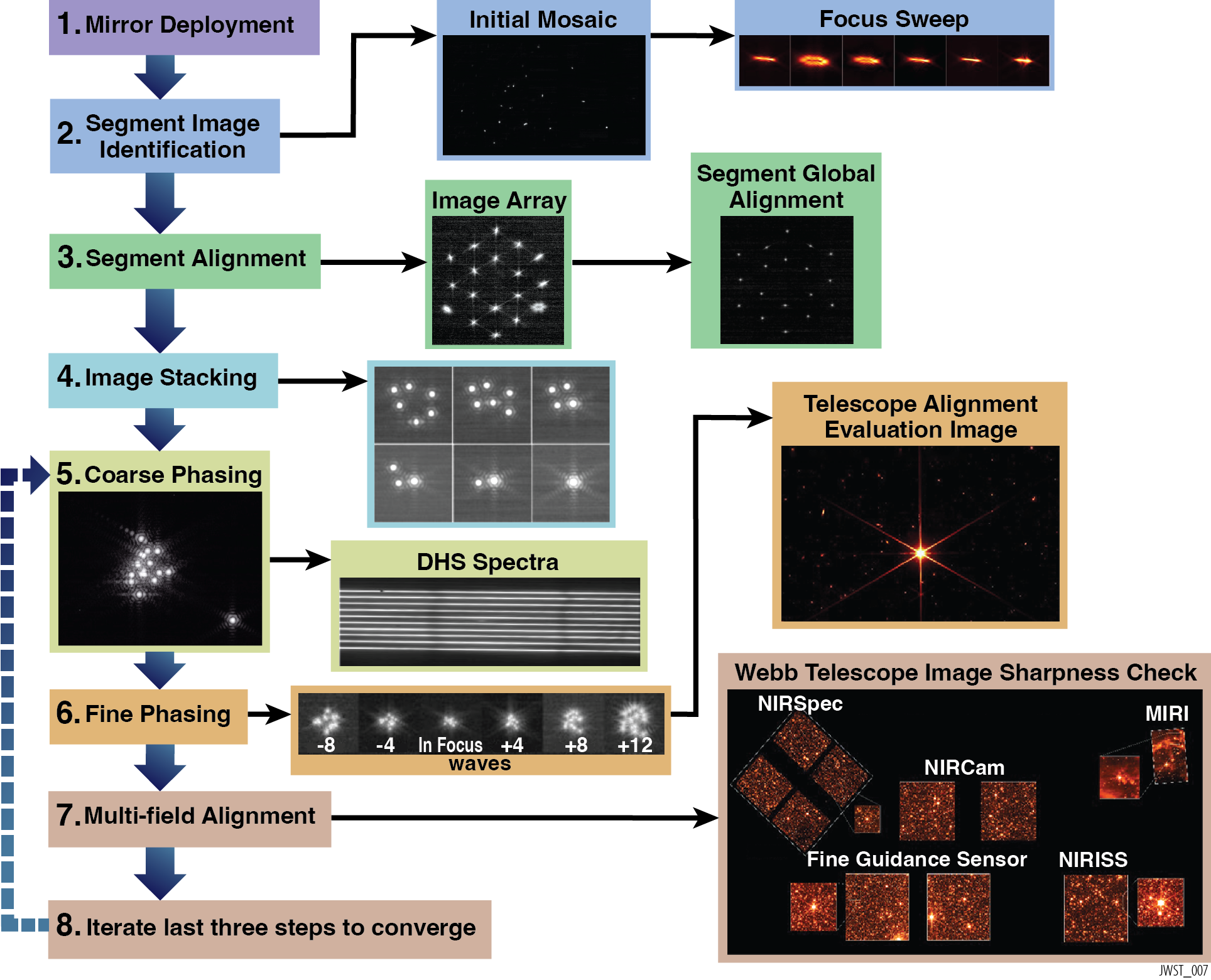}
  \caption{High-level overview of the sequence of events during OTE commissioning, along with examples of flight data. This depiction greatly simplifies a complex process involving hundreds of individual steps and observations.} 
  \label{fig:CommissioningSequence}
\end{figure*}

To prepare the telescope for scientific observation, OTE commissioning activities included the deployment of the mirror segments from their launch restraints, the alignment of the primary and secondary mirror segments, and the achievement of a finely phased telescope. The series of activities was developed and rehearsed over many years and was allocated $\sim$90 days to complete in the planned schedule. For clarity, it is useful to break the whole sequence down into a few major sets of activities: mirror segment deployment, segment-level identification and alignment, co-phasing of the segments, and multi-field alignment. Here, we summarize these activities and describe the actual execution during flight (see Figure~\ref{fig:CommissioningSequence}). The detailed commissioning plan is presented in  \cite{Acton+Dean+Feinberg+etal_2018} and additional details on the as-run activities are presented in \cite{Feinberg+Wolf+Acton+etal_2022} and \cite{Acton_2022SPIE12180E..0UA}. Briefly, the commissioning plan had to accommodate initial PMSA and SM positional errors as large as one millimeter, and progress to achieve fine alignments within a few tens of nanometers. This was achieved using several different forms of wavefront sensing, most of which were iterated multiple times, and which had to be interspersed with early steps of instrument and guider commissioning and focal plane calibrations.

% Commissioning Sequence

% OTE Commissioning Timeline
\begin{deluxetable*}{lcccr}
\tabletypesize{\footnotesize}
\tablecaption{Timeline Summary for OTE Commissioning \label{tab:CommissioningTimeline}}
\tablewidth{0pt}
\tablehead{\colhead{Activity } & \colhead{ Program ID\tablenotemark{a} } & \colhead{ Prelaunch Plan } & \colhead{ Actual } & \colhead{ Delta [days]}}
\startdata
Mirror Deployments              & \textit{n/a} & 2022-01-18 & 2022-01-20 & 2.0\\
Segment Identification          & 1137 & 2022-02-07 & 2022-02-08 & 0.2\\
First Closed Loop Guiding       & 1410 & 2022-02-12 & 2022-02-13 & 0.5\\
Segment Alignment (iteration 1) & 1141 & 2022-02-12 & 2022-02-19 & 6.2\\
Image stacking (iteration 1)    & 1143 & 2022-02-14 & 2022-02-22 & 7.1\\
Coarse Phasing (iteration 1)    & 1147 & 2022-02-21 & 2022-02-28 & 6.8\\
Coarse Multi Field              & 1148 & 2022-02-22  & 2022-03-03 & 8.9\\
Fine Phasing (iteration 1)      & 1155 & 2022-03-01 & 2022-03-08 & 6.4\\
Fine Phasing (iteration 3)      & 1160 & 2022-03-05 & 2022-03-11 & 5.5\\
Multi Field Multi Instrument Sensing 1 & 1166 & 2022-03-10 & 2022-03-20 & 9.7\\
LOS Jitter Measurement          & 1170 & 2022-03-17 & 2022-03-21 & 4.0\\
Multi Field Multi Instrument Sensing 2 & 1465 & 2022-04-06 & 2022-04-19 & 12.8\\
OTE Alignment Complete          & \textit{n/a} & 2022-04-24 & 2022-04-23 & -1.0\\
\enddata
\tablenotetext{a}{APT program IDs, which may be used to retrieve these data, or any other commissioning data, from the MAST archive.}
\tablecomments{Timeline of events corresponding to Figure~\ref{fig:CommissioningSequence}. Overall the OTE commissioning process proceeded remarkably according to preflight plans. The major sources of schedule deviation were complications in early guiding (delayed completion of segment alignment) and a revised, slower plan for MIRI cooldown adopted by the MIRI team (delayed start of the second round of multi instrument sensing.) These were balanced out by the budgeted third round of multi instrument sensing not being necessary, allowing the completion of OTE alignment one day ahead of schedule.}
\end{deluxetable*}

\subsubsection{OTE Deployments}

The major structural deployments of the OTE (DTA, SMSS, and PMBSS wing deployments) all completed successfully and nominally, with no notable issues. 

Subsequent results from telescope commissioning confirmed the precision of these deployments: for instance, the 8~m multi-hinged SMSS deployment placed the SM within 1.5~mm of its nominal position, well within the correction range of the SM actuators. Further, the telescope boresight offset relative to the spacecraft star trackers was found to be 3.4$\arcmin$, nicely consistent with the 3.05$\arcmin$ 1-sigma preflight prediction. Similarly the corrections required to align the PMSAs were small (Table \ref{tab:pmsa_dofs}), with only one segment requiring a corrective move larger than 1 millimeter in position. 

\begin{deluxetable}{lcccc}
\tablecaption{PMSA Correction Magnitudes \label{tab:pmsa_dofs}}
\tablehead{\colhead{Degree of Freedom} & \colhead{Unit} & \colhead{Typical} & \colhead{Maximum} & \colhead{Margin}}
\startdata
Piston & $\mu$m & 145 & 281.7 & 91\%\\
Radial Translation  &  $\mu$m & 450 & 1205.5 & 32\% \\
Clocking & microradian &  370 & 766.0 & 55\% \\
Radius of Curvature & $\mu$m of surface sag & 0.75 & 2.117 & 76\% \\
\enddata
\tablecomments{Typical and maximum corrective moves required to align the PMSAs. OTE deployments initially placed segments typically within a few hundred microns of their intended locations. The margin column gives the unused fraction of the nominal correction range remaining after the maximum correction moves for each degree of freedom.}
\end{deluxetable}

\subsubsection{Mirror Segment Deployment}

\begin{figure}
  \centering
  \includegraphics[width=0.45\textwidth]{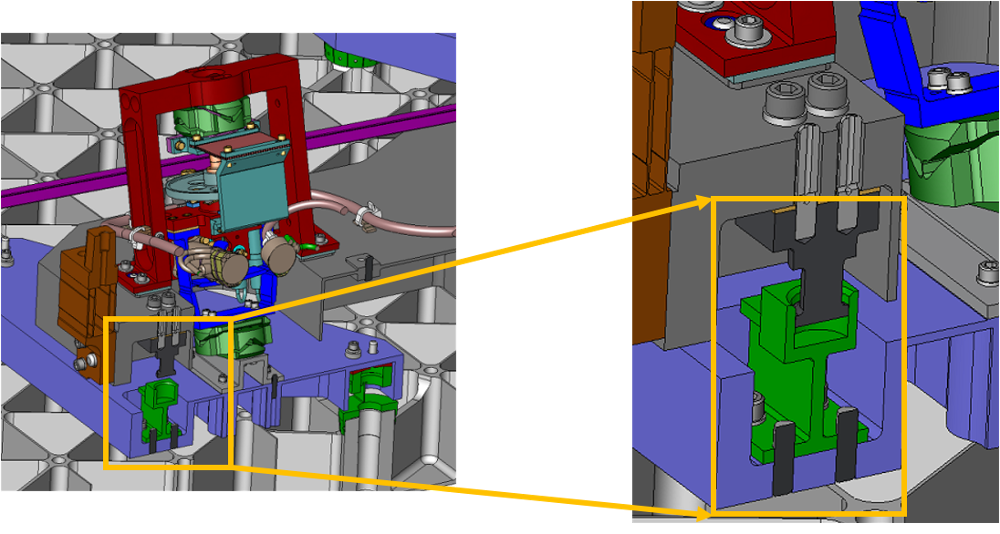}
  \caption{Rendering of the actuator in a stowed configuration, with the snubber inside the launch restraint (in green).} 
  \label{fig:MirrorDeployment}
\end{figure}

For launch, all mirror segments were stowed in launch restraints in order to limit lateral displacements during launch and ascent (see Figure~\ref{fig:MirrorDeployment}). To begin to align the PMSAs and SMA, the segments therefore had to be released from their launch restraint, a pure piston move of $\sim$12.5~mm. 

% Mirror Deployment

The deployment sequence of the PMSA and SMA was carefully designed to verify the actuator stepper motor aliveness and responsiveness, as well as confirm proper management of the segment envelope and workspace boundaries. The sequence was tested on the ground on multiple occasions (e.g., OTIS testing at JSC). As such, stepper motors on segment A1 only were first incrementally commanded to move 1 step, 1 revolution, 10~$\mu$m, 150~$\mu$m, and 340~$\mu$m. The remaining segments (except A3 and A6, see below) were then commanded the same sequence of steps, followed by 1-mm increments until all the segments were fully deployed to 12.5~mm. Early in these deployments, some LVDT sensor readings did not show as smooth a progression as expected, which led to additional small ‘flinch’ moves to verify all actuators were moving. Once past the first few millimeters and out of the launch restraints, the LVDTs showed the expected linear response. The initial non-linearity was interpreted as due to friction of surface contacts with the launch restraints, which, however, posed no problem to the deployments.  

Segments A3 and A6 were deployed separately and last as a result of a faulty LVDT on one actuator each (as noted above in Section~\ref{sec:otis_ambient}). Although the sequence of moves was identical to that of the other segments, the LVDT readings, which provide a coarse direct measurement of actuator length, had to be calibrated as the sensors cooled down. As a result, they were deployed separately, without incident.

Finally, once at their deployed positions, the segments were commanded to their intended nominal positions, based on ground alignment test results corrected for 0-g via modeling. Mirror deployments completed successfully and with no major issues. 

\subsubsection{Segment-level Identification \& Alignment}

\begin{figure*}
  \centering
  \includegraphics[width=\textwidth]{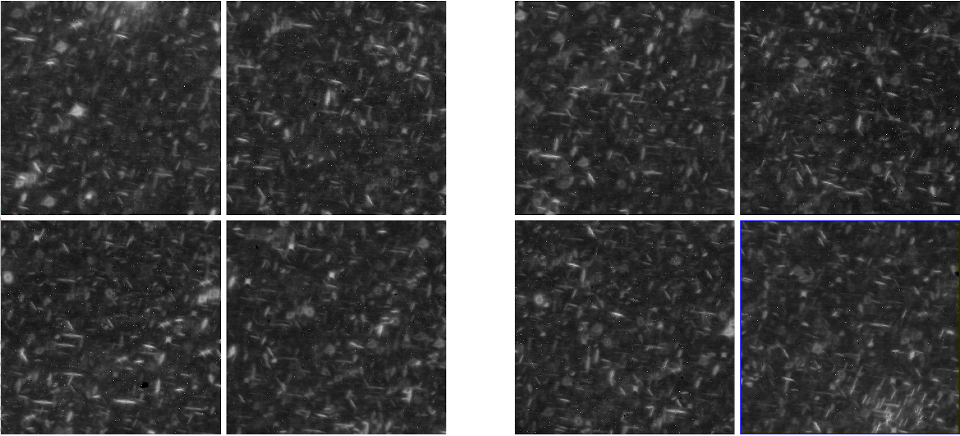}
  \caption{First on-sky image using JWST and NIRCam, targeting a region of the Large Magellanic Cloud, obtained on 2022 February 2. Because the mirror segments are not phased, each star is duplicated 18 times. Two copies of the core of globular cluster NGC 1916 can be seen at upper left and lower right.}
  \label{fig:SegmentIdentification}
\end{figure*}

The next series of activities aimed at finding, identifying, and re-arranging the images produced by each of the mirror segments. The deployed but not-yet-aligned segments were each acting as its own aberrated $\sim$1.3~m telescope. When first light on NIRCam was obtained on February 2$^{nd}$ 2022, all celestial objects were indeed duplicated 18 times (see Figure~\ref{fig:SegmentIdentification}). 

% Segment Identification

\begin{figure}
  \centering
  \includegraphics[width=0.45\textwidth]{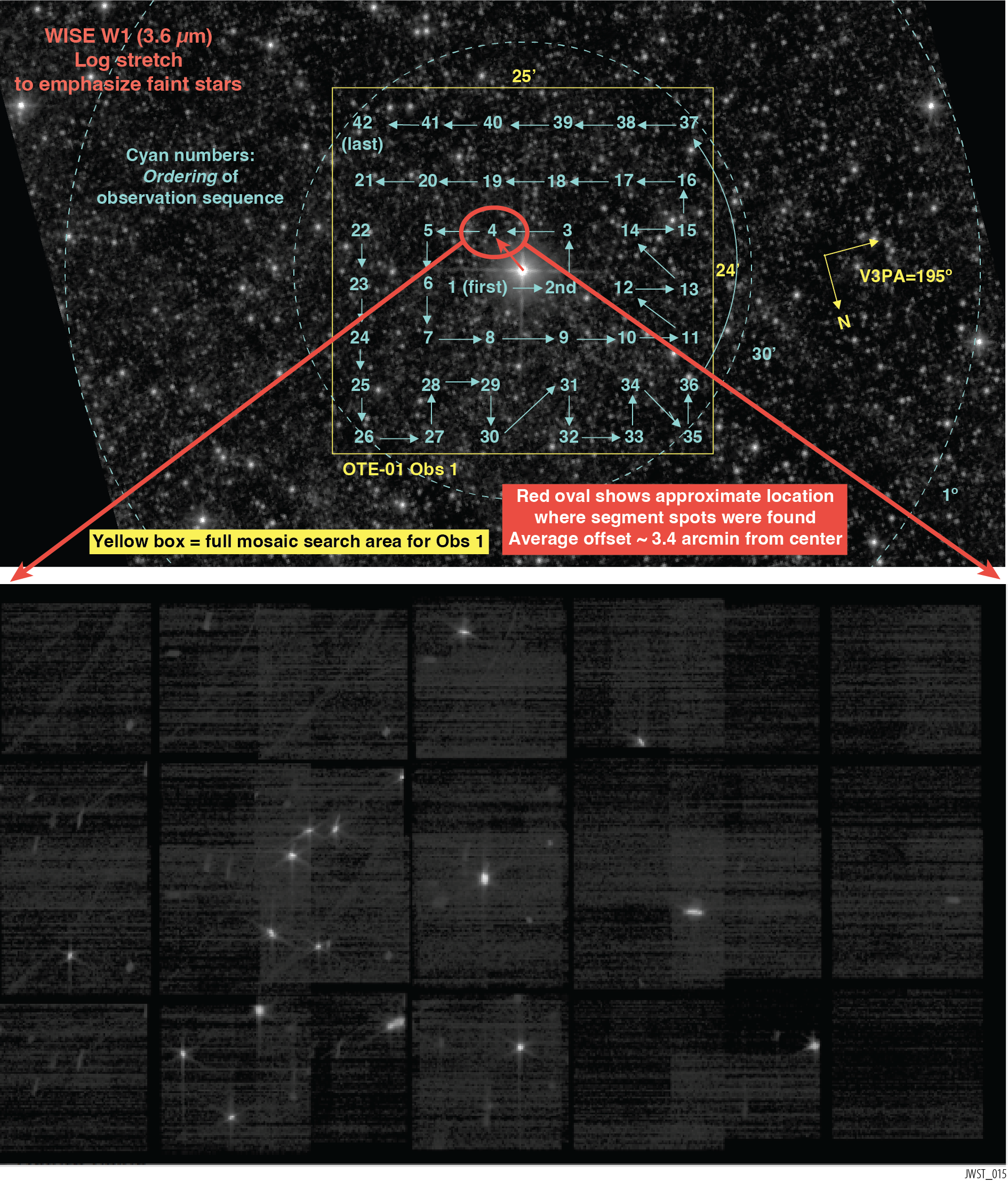}
  \caption{Top: The search sequence of NIRCam observations centered on the bright target HD~84406. Bottom: The inset shows a subset of the large mosaic from flight. All 18 images of the target star were found clustered together in a relatively small area, due to good deployments precision. Also visible are some residual after images of stars and curved trails from illumination during slews to different pointings, due to increased persistence from higher-than-nominal detector temperatures at the time of these observations.} 
  \label{fig:NIRCamMosaic}
\end{figure}

%(HD 84406, which has a K band magnitude of $\sim$4.7)
To identify an image of each segment, an isolated bright star was observed. Budgeted pre-flight uncertainties for mirror segment initial deployments, as well as in the initial telescope boresight offset (see Section~\ref{subsubsec:CalTel2Spacecraft}), predicted that segments might be scattered by up to $\sim$15$\arcmin$. As a result, the target star was selected to have no similar-brightness neighbors within such a distance, and a large half-degree-diameter mosaic around the target star was generated, taking around 25 hours to complete. Figure~\ref{fig:NIRCamMosaic} shows a cartoon of the planned mosaic, overlaid on top of a catalog sky image around the target. A subset of the flight data where the segment images were all found is also shown. The segment images were found within $\sim$3.4$\arcmin$ of the nominal target location on average and the segment image scatter was of similar magnitude, both better than requirements and expectations.  The simplicity of the plan for this initial mosaic step proved beneficial to accommodate larger-than-expected coarse pointing uncertainties at this time (prior to ACS tuning and optimization), and increased levels of detector persistence due to operating NIRCam well before it had fully cooled to its nominal temperature.

% Image Mosaic

Following the initial mosaic observation, a secondary mirror focus sweep was performed in order to measure and move the SM to a best focus position. This improved image quality and enabled guiding later on during OTE commissioning (see Section~\ref{subsubsec:GuidingCommissioning}). The analysis led to an SM move of -427~$\mu$m (i.e., away from the primary mirror).

Next, each mirror segment's image in the initial mosaic was identified by sequentially tilting each mirror. Once identified, segments were commanded to form the pre-defined image array shown in Figure~\ref{fig:CommissioningSequence}. 

In the hexagonal array configuration, all segment images could be observed on one NIRCam detector at once and segment-level aberrations could be addressed as part of the segment global alignment activities. To do so, the SM was moved away from its nominal best focus by $\pm$400~$\mu$m to collect focus-diverse imagery for the purpose of phase retrieval analysis. As a result of this analysis, the SM was corrected in X- and Y-translation by 0.94 and 1.06~mm, respectively. Focus corrections were applied to each segment, along with corrections to the radius-of-curvature actuators of two PMSAs. 

\begin{figure*}[ht]
  \centering
  \includegraphics[width=0.8\textwidth]{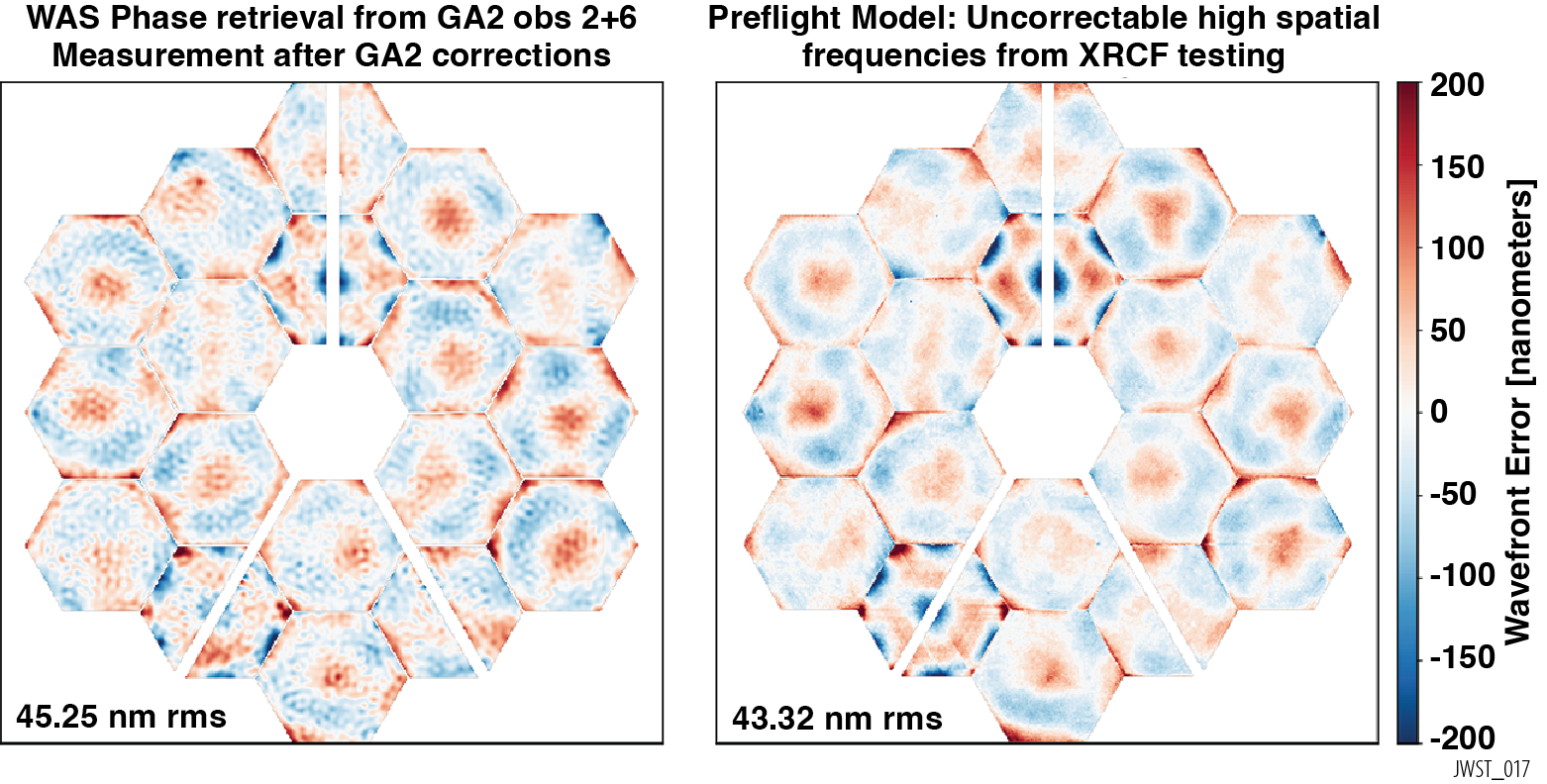}
  \caption{On-orbit measured primary mirror segment wavefront errors following global alignment 2 (left) as compared to the ground-test measurements at the XRCF with model backouts for 0-g gravity (right). The consistency of these two datasets provides a striking in-flight validation of the OTE development program. It demonstrates the segments were fabricated to the correct optical prescriptions, the 0-g gravity backouts were correct, there was no measurable thermal distortion during launch and ascent, the segments were deployed to within the correction range of their actuators, and the WFSC processes worked to sense and correct the initial misalignments. Note that the RMS WFE values labeled are for the primary mirror segments only, excluding all the other optical surfaces which contribute to higher WFE for the observatory as a whole.
  }
  \label{fig:segment_wavefronts}
\end{figure*}

During a later stage of commissioning, a second iteration of Global Alignment was executed as part of the iterative alignment approach. At that time, additional small clocking, radial translations, and radius of curvature corrections were applied to the PMSAs to correct astigmatism and power. At this point in the OTE phasing process, the observed wavefront errors achieved excellent agreement with ground measurements and preflight modeling of the higher spatial frequency mirror maps (Figure~\ref{fig:segment_wavefronts}).

% WFE maps

\subsubsection{Co-Phasing the Segments}

Following each instance of Global Alignment, a sequence of image stacking was executed to position segment images on top of each other (see Figure~\ref{fig:CommissioningSequence}). However, this stacking does not mean the light paths from each segment are in phase with one another, so Dispersed Hartmann Sensing measurements (Coarse Phasing) were executed to establish phase errors as a function of wavelength in order to measure pairwise segment edge heights and derive an overall piston correction to all of the PMSAs.  Three iterations of this coarse phasing sufficed to bring the PMSA piston offsets to less than 1~$\mu$m, where fine phasing with NIRCam weak lenses could finalize the OTE alignment. 

Following these PMSA piston corrections, a fine phasing activity took place where the NIRCam weak lenses were used to collect focus-diverse measurements for phase retrieval analysis (see Figure~\ref{fig:CommissioningSequence}). The results of such measurements were then used to better (re-)stack the segment images as well as correct residual piston offsets between segments. 

On March 11 2022, the fine alignment process completed, yielding a telescope aligned to roughly 50~nm RMS as seen at the fiducial field point on NIRCam A3. Following this, a mosaic observation was carried out around the alignment star. This multi-purpose observation tested science-like dithered and mosaiced observations for the first time, confirmed excellent PSF quality over all of NIRCam's field of view and across NIRCam's full wavelength range, provided early measurements of observatory backgrounds, and yielded an early glimpse of JWST's sensitivity to the high redshift universe. 

\subsubsection{Multi-Field Alignment}

Aligning the telescope to only one field can lead to degenerate solutions, where PMSA and SM misalignments balance each other out. Multi-field measurements are therefore required in order to achieve optimal optical performances across all the science instruments. 

Two types of multi-field sensing were carried out: a NIRCam-only multi-field activity (using only Module A) and two instances of multi-field, multi-instrument activities. The NIRCam-only alignment activity was aimed at initially removing most of the PMSA-to-SM misalignment using the unstacked segments as a Hartmann sensor, analyzed with a centroid-based approach to measure the field dependence of aberrations.  This ``coarse'' multi-field sensing proved to work exceptionally well. As a result of this analysis, the SM was moved in translation in X and Y (-210$\mu$m, 420$\mu$m) as well as tilt in X and Y (-550~$\mu$rad, 34~$\mu$rad). The SM correction was also accompanied by compensating PMSA moves in order to maintain the hexagonal array configuration.

\begin{figure*}
  \centering
%  \vspace{-0.7in}
  %\hspace{-1.5in}
  \includegraphics[width=0.85\textwidth]{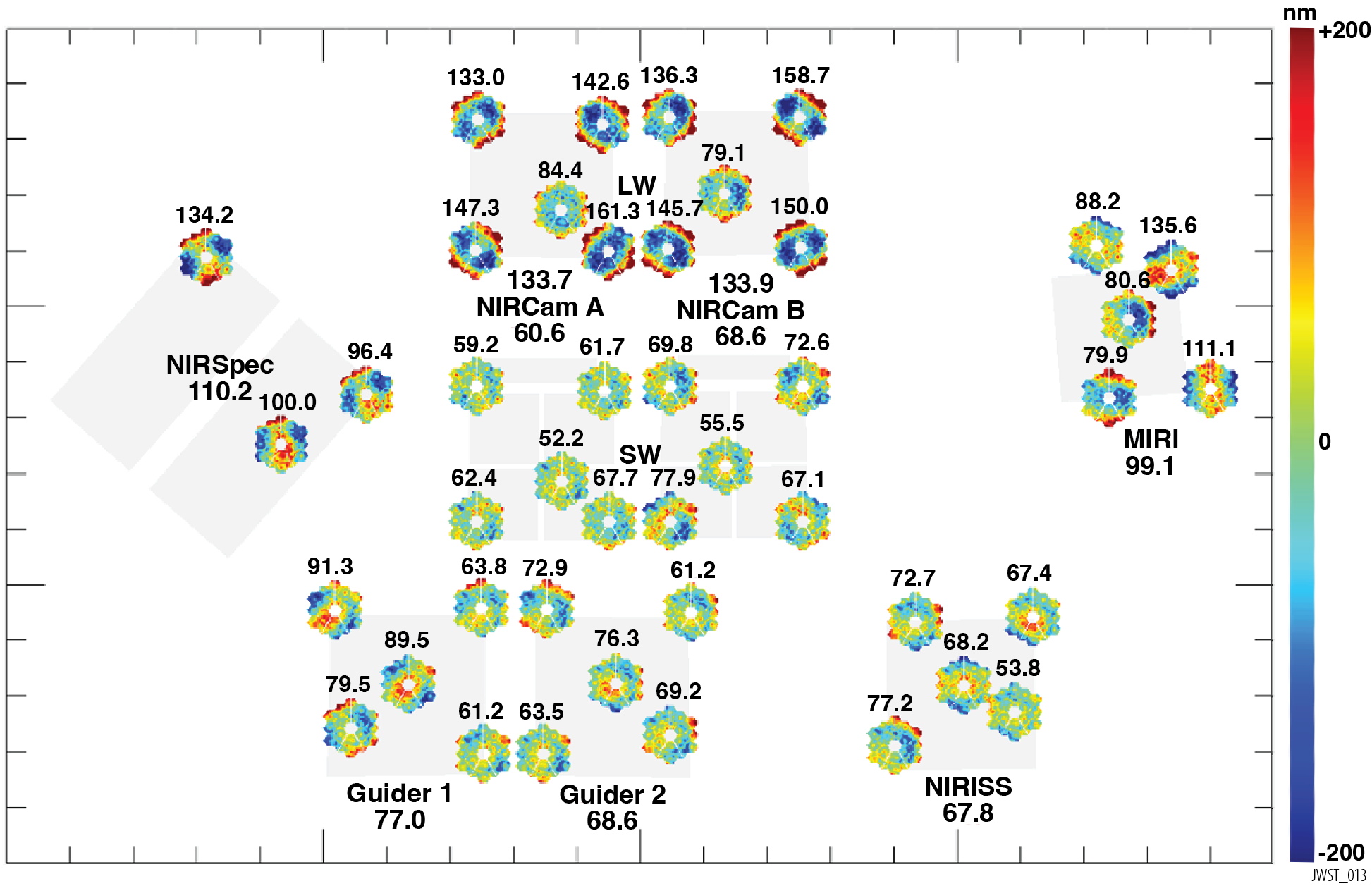}
%   \vspace{-0.9in}
  \caption{Measured multi-field, multi-instrument wavefront residuals, here showing the end-to-end observatory (i.e. OTE plus instruments) static wavefront residuals at the end of telescope alignment. The variation between field positions seen here is due mostly to the inherent optical performance of the instruments, outside of the ability of OTE adjustments to correct. The final adjustment after multi-instrument sensing was only a small focus shift of the secondary which brought the science instruments into good average focus and positioned the global focus closer to MIRI. All instruments maintained precise confocality from ground to space, such that it was unnecessary perform focus adjustments on the individual instruments.} 
  \label{fig:MIMF}
\end{figure*}

Finally, multi-field, multi-instrument measurements were made in order to assess the field dependence over the whole field of view, this time using focus diversity provided by moving the SM in piston by $\pm$100~$\mu$m. Two instances of this activity were executed since, as expected, MIRI had not yet reached its operational temperature at the time of the first measurements. In both instances, the results indicated no significant correctable field-dependent aberrations, in other words the initial NIRCam-only multi-field activity had corrected the telescope’s field dependent aberrations. Only a small SM focus correction with minimal wavefront error gain was applied, mostly to balance the relative focus terms of the science instruments (see Figure~\ref{fig:MIMF}). The observatory was fully aligned on April 23 2022, and the commissioning of the science instruments continued thereafter. 

From that point, the operations team entered a new stage of conducting routine OTE maintenance, which will be ongoing throughout the mission, discussed further in Section~\ref{sec:ScienceEra}.

% MIMF

\subsection{Guiding and Line-of-Sight Pointing} \label{subsection:Guiding&LOSPointing}

To support all but the very earliest mirror alignment activities described in Section~\ref{subsec:OTECommissioningActivities}, the closed-loop FGS guiding mode first needed to be commissioned and its performance established. See Menzel PASP Observatory for a more detailed discussion of the Attitude Control System.

\subsubsection{Guiding during Commissioning} \label{subsubsec:GuidingCommissioning}

JWST uses star trackers, rate sensors, reaction wheels, and a fine steering mirror to achieve a coarse pointing. To transition into guiding, one of two FGSs will attempt to identify the intended guide star, whose position and fluxes are normally provided by the Guide Star Catalog. The coarse position error of the guide star as seen by FGS is fed back to ACS for correction. The guidestar will usually then be moved to a pre-computed ``science'' location in the FGS field of view, after which closed-loop guiding will be attempted and typically engaged. During closed-loop guiding, FGS measures centroids every 0.064~s and reports the position to ACS, which then commands the fine steering mirror to move in order to maintain the guide star at the appropriate position on the FGS detector.

Early in OTE commissioning, when the OTE was still providing 18 images for every star in the field of view (including on FGS), guiding operations had to be modified. Per our plan, FGS used one of the segment images to guide on, along with reference segment images. This was done by overriding the guide star selection system to account for the segment position offsets and flux differences compared to the guide star’s catalog position and flux. Using this approach, closed-loop guiding was successfully demonstrated during the FGS LOS Initialization Activity (PID 1410).

Later in commissioning, guiding operations became increasingly routine; once the PMSAs were stacked into a single PSF, the guide and reference stars’ selection and locations could be automatically provided by the ground system using the operational catalog and only the associated stellar fluxes had to be overridden. Once the PSF was phased, the fluxes as well were being supplied by the system. By the end of OTE commissioning, guiding required no special intervention.

The majority of instances of closed-loop guiding during OTE commissioning were successful, although some failures do occur for a variety of reasons, including bad pixels or mis-cataloged guide stars. Nevertheless, success rates have risen continuously throughout commissioning and into science operations with now $>$~95\% of planned visits being successfully executed. When guiding success is achieved (i.e., the intended guidestar is identified, acquired, and tracked with a settled closed loop), then the image stability of that pointing becomes the performance metric of interest.

\subsubsection{Characterizing Line of Sight Stability during Commissioning} \label{subsubsec:LOSstability}

Special commissioning tests were included in the baseline plan to obtain the power spectrum of the line-of-sight (LOS) jitter and, in particular, to assess the contribution of vibrations caused by the MIRI cryocooler (CC).

\begin{figure*}
  \centering
  \includegraphics[width=0.38\textwidth]{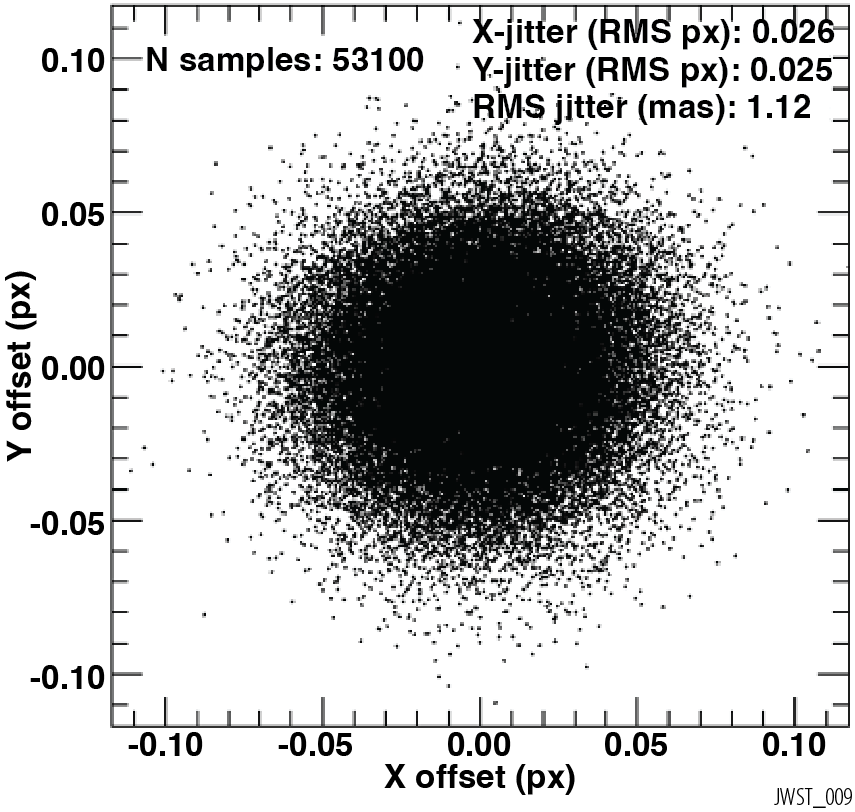}\hspace{0.5in} \includegraphics[width=0.5\textwidth]{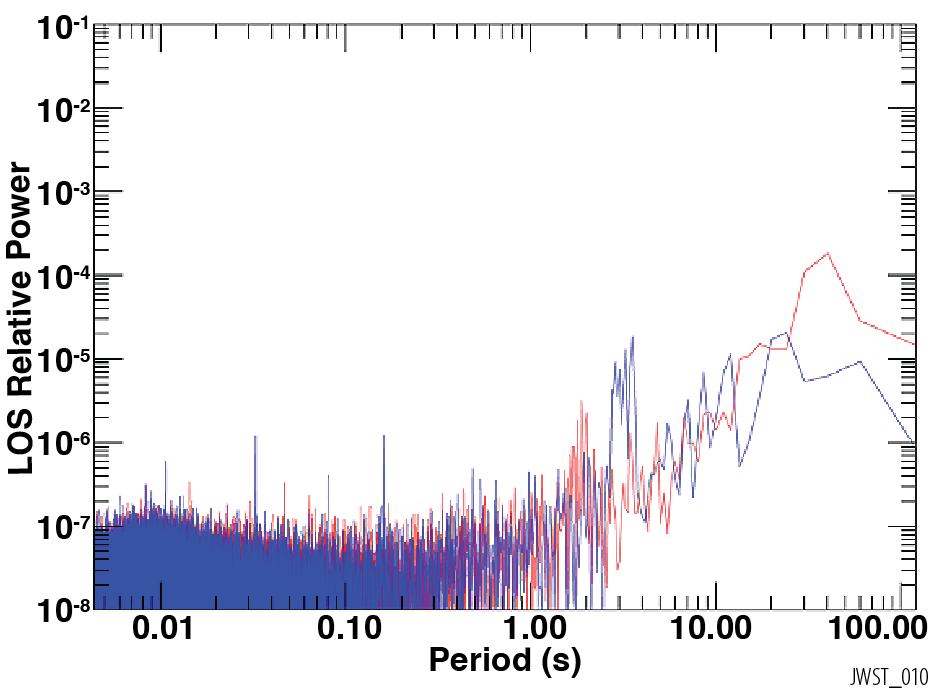}
  \caption{Left: Line-of-sight jitter distribution sampling every 2~ms over a 120~s interval. This ``jitter ball'' shows well-behaved and very small variations in pointing using fine guidance control, with RMS variation only 1.1~mas. Right: The power spectrum from the high-frequency jitter measurement.} 
  \label{fig:jitter} \label{fig:psd}
\end{figure*}

The dedicated test to probe excitations from vibrations in the LOS data showed no evidence of significant contributors and has revealed excellent stability performance at $\sim$1 mas RMS radial, very close to the measurement noise floor and significantly better than expectations and the 7~mas requirement. The typical results from the first commissioning LOS jitter analysis (PID 1163, observation 2, from \citealt{Hartig_2022}) are shown in Figure~\ref{fig:jitter}. In this case, jitter was measured at 1.04~mas RMS (radial). The same tests also revealed no need to tune the MIRI cryocooler’s pulse frequency which remains at its initial settings. 

The analyses, which continue to be regularly performed as part of the 2-day routine maintenance program (Section~\ref{sec:ScienceEra}), have consistently shown jitter levels around 1~mas radial (moderately correlated with guidestar brightness). The analysis is quite sensitive and has revealed low frequency, low power oscillations at $\sim$0.3 and 0.04~Hz. The 0.3~Hz feature has been attributed to bending modes at the 1~Hz isolator at the SC-to-OTE interface, whereas the 0.04~Hz feature has been shown to vary over time and might be attributed to fuel slosh. During the commissioning period there were no clear indications of any LOS jitter response to the reaction wheel assemblies. However during the first months of science operations, a handful of measurements have shown minor LOS jitter contributions which appear correlated with certain speeds of reaction wheel assembly \#6; a resonance appears to be excited in the vicinity of 16--17 Hz. 

% Jitter 

\subsection{Focal Plane Calibration} \label{subsection:FPCals}

In addition to guiding (Section~\ref{subsection:Guiding&LOSPointing}), sufficiently accurate target placement at each of the SIs was needed to support late OTE commissioning activities. This involved two essential related areas of calibration: the first was the determination of the principal coordinate frame of FGS1 with respect to the spacecraft’s, expressed as a direction cosine matrix and stored on-board for use by the ACS. The second was the determination and calibration of the SI fields of view relative to FGS1, expressed in a convention similar to Euler angles and stored in the SOC ground system for determining on-sky pointings. These activities were performed in close coordination with WFS$\&$C as part of the integrated flow through OTE commissioning.

\subsubsection{Calibrating the Telescope to Spacecraft} \label{subsubsec:CalTel2Spacecraft}

A key design feature of JWST is an OTE that is well separated and isolated, thermally and mechanically, from the spacecraft by the Deployable Tower Assembly (DTA). Uncertainties in the DTA deployment and other contributions were expected to produce initial errors in the nominal alignment of the OTE V-frame with respect to the spacecraft’s fundamental coordinate system, called the J-frame, of $\sim$10–15$\arcmin$.

The slew of the observatory to its intended field is controlled by the Star Tracker Assemblies (STAs), which reside on the spacecraft side and are calibrated to the J-frame. However, the FGSs must then be able to acquire and identify the intended guide star, and they reside on the OTE side. Capturing the alignment between FGS and ACS requires the updating of the FGS-to-J direction cosine matrix (DCM). This matrix is updated to account for deployment uncertainties, as stated above, as well as changes in telescope boresight, which occur every time the SM is moved in translation or tilt (e.g. Global Alignment, MIMF).

While the initial error in the OTE-to-spacecraft alignment was expected to be close to 10$\arcmin$, the misalignment was found to be only $\sim$3.4$\arcmin$ from the nominal ground values (see Figure~\ref{fig:NIRCamMosaic}). The FGS-to-J DCM was subsequently manually updated during commissioning after taking observations of the sky with an SI or FGS and expressing that celestial pointing in terms of the FGS1 frame, while obtaining from the ACS/STAs the contemporaneous mapping of sky to the J-frame, thus providing the information to relate FGS1 to J-frame. This operation was also successfully performed during subsequent commissioning activities to maintain the alignment. 

During the science mission, this relationship will continue to be dynamic at levels much lower than seen in commissioning (i.e. a few arcsec), and the ACS will autonomously update this calibration based on observed FGS guidestar location error.

\subsubsection{Calibrating the SIs to FGS}

Using FGS1 to define the relationship of the OTE to the spacecraft implies that it is also the reference for the OTE-based frame to which the SI fields of view are calibrated. This ``V-frame'' was defined nominally as a conventional 3-axis coordinate system aligned with the OTE principal mechanical axes, having V1 pointing out along the Cassegrain axis of symmetry (\ref{fig:ObservatoryArchitecture}). In flight, however, the V-frame is used to specify SI and FGS fields and various fiducial field points used for science targeting. So, in this application, it is treated spherically as angles, with the axes V2 and V3 corresponding to ``field angles'' within the OTE field.

In this scheme, the FGS1 field location and orientation with respect to V2,V3 is fixed, and on-sky astrometric calibrations that determine the SIs’ and FGS2’s fields relative to FGS1 in essence establish their locations, orientations, and higher order distortions with respect to the V2,V3 field angles. This astrometric calibration scheme, its tools and products, are thoroughly treated by \cite{Sahlmann_2019a}. 

These calibrations are required for successful target placement and were performed during OTE commissioning to (1) determine the post-launch changes to the ground-determined relationships and (2) update this knowledge to ensure successful multi-SI wavefront measurements. The commissioning team used a specially calibrated $\sim$15$\arcmin$ astrometric region of the LMC for this purpose \citep{Sahlmann_2019b}.

The first measurements of SI relative locations showed the ISIM to be stable, with ground-to-flight evolution in the V2,V3 field angles to be at or below the $\sim1''$, and orientation changes of the SI fields $<$ 1$\arcmin$. Although precise scales and distortion calibrations fall into the SI activities and continue into the science cycles, basic instrument scales were tentatively measured during these OTE commissioning activities, and were found to be $<$ 0.15$\%$ different from ground measurements. For comparison, Hubble’s SIs through the generations typically saw 1-$2''$ of V2,V3 shifts and 0.2-1.0$\%$ scale change from ground to flight.

\subsection{Optical Performance at the End of Commissioning}

The optical performance as measured at the end of commissioning is better than the requirement values at the system-level and for the most part, better than the sub-system allocations as well. A summary of the optical performance measurements and predictions is presented in this section. A separate optical discussion is presented in the paper on JWST Science Performance in this volume by Rigby PASP Science. A detailed discussion of the telescope’s optical performance against requirements at the end of commissioning is reported in \cite{Knight+Lightsey_2022}.

The total wavefront error combines the observed static wavefront error with the dynamic stability and image motion terms, which is corrected when the NIRCam A field point exceeds 80~nm rms. This means the end-to-end, telescope and NIRCam SW has diffraction-limited image quality at $\sim$1.1~$\mu$m. This is significantly better than the 150~nm~RMS requirement optical error budget total that enables diffraction limited image quality (approximated by $\lambda$/14) at 2~$\mu$m. 

\subsubsection{Observed Wavefront Errors} \label{subsubsec:ObsWFE}

Table~\ref{tab:WFETable} summarizes the end-to-end, referred to as `observatory', static wavefront errors measured at the end of commissioning. The static wavefront errors are well below their allocations in all channels, at all field points.

% WFE Table 
\begin{deluxetable}{cc}
\tabletypesize{\scriptsize}
\tablecaption{WFE Table  \label{tab:WFETable}}
\tablewidth{0pt}
\tablehead{ 
\colhead{Science Field} & 
\colhead{Static WFE} 
}
\startdata
NIRCam A SW & 61$\pm$8 \\
NIRCam B SW & 69$\pm$11 \\
NIRCam A LW & 134$\pm$38 \\
NIRCam B LW & 134$\pm$39 \\
NIRISS & 68$\pm$12 \\ 
FGS 1 & 77$\pm$15 \\ 
FGS 2 & 69$\pm$8 \\ 
MIRI & 99$\pm$28 \\ 
NIRSpec & 110$\pm$20 \\ 
\enddata
\tablecomments{The static observatory wavefront error measurements as measured at the end of telescope alignment in May 2022. The values reported are the average across the points measured across the science field, with the plus and minus values reporting the peak to valley variations. The total observatory WFE combines this static term with the dynamic WFE stability and image motion, typically $\sim$13~nm rms equivalent, via RSS sum. Units are nanometers rms.}
\end{deluxetable}

\subsubsection{Area \& Throughput}

JWST’s unobscured collecting area was measured using the NIRCam pupil imaging lens to be 25.44~m$^{2}$, exceeding its requirement of 25 m$^{2}$. The telescope’s wavelength-dependent transmission ranges from 0.786 at 0.8~$\mu$m to 0.933 at 28~$\mu$m, again better than requirements at each wavelength. The transmission values were determined from final pre-flight measurements of mirror witness samples, combined with NIRCam grism measurements confirming the absence of detectable ice deposits. 

The product of the above observed values for OTE area and transmission was also projected to end of life using modeled degradation of the optics. This places the effective area $\times$ transmission value of 19.58 m$^{2}$ at 0.8~$\mu$m and 23.18 m$^{2}$ at 20~$\mu$m, compared to the OTE requirements of 15.37 m$^{2}$ and 22.00 m$^{2}$, respectively.

\subsubsection{Vignetting}

Some of the OTE commissioning activities described earlier obtained data that also supported a secondary goal of probing various types of vignetting. Analyses of these data show no indication of any field-of-view cropping, unexpected OTE structure incursion, or pupil vignetting. Establishing that the telescope was unobstructed (except for secondary mirror support structures) fulfilled a mission-level requirement and was an exit criterion for OTE commissioning.

\subsubsection{Thermal Stability} \label{subsubsec:ThermalStability}

A dedicated thermal stability test was carried out following the telescope alignment in order to characterize the wavefront stability and image motion on various timescales following a large, stressing thermal slew (i.e. early May 2022; PID 1445 and 1446). This activity started by performing a 4-day thermal soak at the hot (sun-normal) attitude and making baseline measurements. Then, the telescope was slewed to the cold attitude where continuous wavefront measurements were made for the first 24 hours and then every $\sim$8 hours for the following 7 days. The thermal stability test confirmed three predicted wavefront drifts that were bounded by the modeling predictions: short-timescale (2-4 min) oscillations from IEC panel heater cycling, medium-timescale ($\sim$1 hr) drift from soft-structure induced thermal distortion, and long-timescale ($\sim$1.5 day) drift from the composite backplane induced thermal distortion. These drifts are reported in Table~\ref{tab:StabilityPerformance}. Temperature sensors on the telescope were also monitored during this test and confirmed that the temperature changes observed were within the noise of the temperature sensors ($<$ 40~mK).  

% Instability Table
\begin{deluxetable}{cccc}
\tabletypesize{\scriptsize}
\tablecaption{Stability Performance  \label{tab:StabilityPerformance}}
\tablewidth{0pt}
\tablehead{ 
\colhead{Contributor} & 
\colhead{Predicted Amplitude} & 
\colhead{Measured Amplitude} &
\colhead{Measured Response} \vspace{-0.2cm}\\
\colhead{} & 
\colhead{WFE (nm RMS)} & 
\colhead{WFE (nm RMS)} &
\colhead{}
}
\startdata
IEC Heater Cycling & 3.5 & 2.5 & 224~s period oscillation \\ 
Frill \& PMSA Closeout & 9 & 4.45$\pm$0.19 & 0.77~hr time constant \\ 
Thermal Distortion & 17 & 17.94$\pm$0.39 & 33.94~hr time constant\\
\enddata
\end{deluxetable}

Some science observation modes are, however, sensitive to these levels of WFE drift. It is also important to note that the worst-case delta-T induced as part of this thermal slew test would rarely be realized during normal science. In practice, science pointings across the sky are subject to much smoother and smaller temperature changes.

Pointing stability immediately following the thermal slew was measured when slewing from hot-to-cold and from cold back to hot. As discussed in Section~\ref{subsec:DesignImplementation}, thermal distortion at the star tracker could result in uncorrected roll about the location of the fine guide star on the fine guidance sensor. The roll about the guide star was measured at the NIRCam field location to be 0.0265 mas/hr in translation and comparable to a measured radial displacement from the star of -0.0230 mas/hr, well below the allocation of 6.3 mas. 

During commissioning, there were many instances of sudden positional changes in one or more mirror segments, referred to as `tilt events' (see Section~\ref{subsubsec:TiltEvents}). The largest of these produce brief violations of the nominal stability values reported in Table~\ref{tab:RequirementSpace}. These positional changes are typically very small but detectable. These tilt events are generally ascribed to strain release within the OTE structures following cooldown to cryogenic temperatures, although the sources of the tilt events is not fully understood.  The frequency and magnitude of these events appear to be slowly declining.

\section{Science Era Characterization} \label{sec:ScienceEra}

The maintenance and trending of the OTE in the science mission era officially started in mid-July 2022 alongside the start of the Cycle 1 science program. The telescope alignment state will be monitored and corrections will be made as needed. Additionally, trending will be carried out across the telescope performance. The telescope state will be made available such that it can be used as part of the science analysis.

\subsection{Wavefront Routine Maintenance Operations Concept}

\begin{figure}
  \centering
  \includegraphics[width=0.48\textwidth]{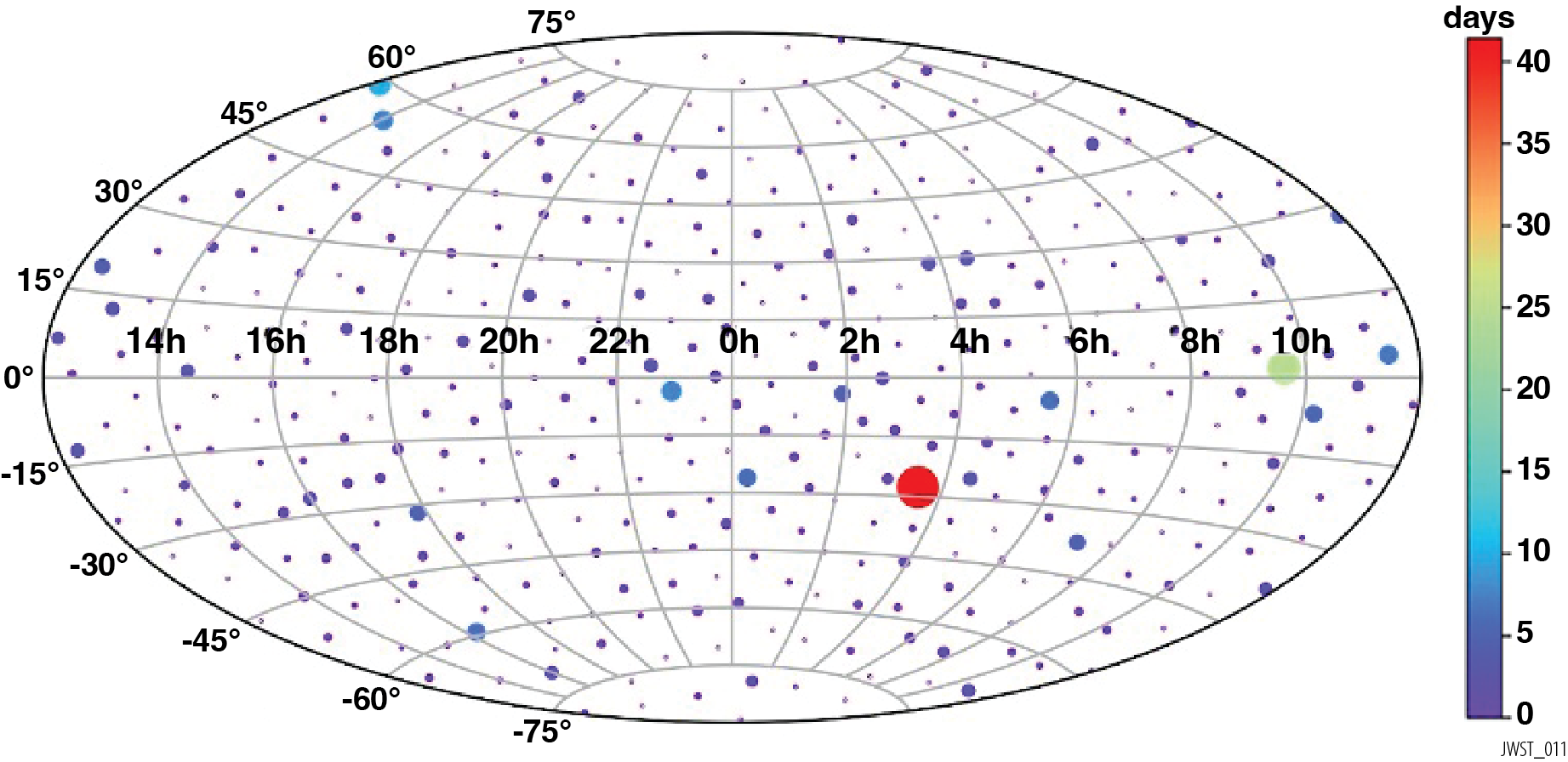}
  \caption{Full sky map of the Cycle~1 wavefront routine maintenance targets, which include 400 targets that were vetted with binary and isolation criteria. The color and size of each dot represent the sum of the time (in days) of Cycle~1 science visits that are closest to that location. Small open purple targets have no nearby science visit in the Cycle~1 plan. The large red circle is $\sim$41 days of observations in GOODS-South and the green circle to the right is the $\sim$23 days of observations in the COSMOS fields. The average distance of a science pointing from a WFSC target is 4.7$\degr$.} 
  \label{fig:Cycle1Pointings}
\end{figure}

The baseline science operations concept for OTE maintenance uses wavefront sensing and control observations to maintain the optical alignment near its optimal state. Wavefront sensing observations are scheduled approximately every 2 days and make use of the NIRCam $\pm$8 wave weak lens pair that is best matched to sensing aberrations at the nominal sensing field point (on detector A3). These observations use bright K$\sim$7 target stars to minimize the exposure times and the exact target is automatically drawn from a pool of targets evenly distributed on the sky in order to minimize slew times between science observations (see Figure~\ref{fig:Cycle1Pointings}). Also included in those routine observations are line-of-sight jitter measurement observations, which use the same target as the weak lens observations and take about two minutes of NIRCam 8$\times$8 data (see Section~\ref{subsubsec:LOSstability}). In total, wavefront sensing observations take about 15 minutes of data, not including slews and overheads (see PID 2586, 2724, 2725, and 2726).

Additionally, pupil images are also collected on a quarterly basis in order to monitor the state of the primary mirror and, in particular, identify and characterize features due to micrometeoroid degradation (see Section~\ref{subsubsec:micrometeoroids} for more details). These observations take about 10~min of NIRCam science time (see PID 2751).

% Cycle 1 Pointings

The sensing data are automatically analyzed on the ground using phase retrieval algorithms (see \citealt{Acton_2022SPIE12180E..0UA}) to assess the state of the telescope and determine if a PMSA control (or correction) is needed. The baseline frequency of such correction was expected to be no more than once every 14 days, which was predicated on the expectation that even worst-case thermal slews would not drift the telescope wavefront error beyond the allocated values. 

Wavefront corrections are typically scheduled when (1) Observatory-level wavefront error exceeds 80~nm RMS or (2) segment tip/tilt correction is larger than 0.05~$\mu$rad. Recall that the observatory-level WFE requirement at NIRCam is 150~nm RMS, so the criterion used here is much tighter thanks to the exceptional in-flight performances of the JWST OTE. Also, these criteria were defined so as to not chase any thermally-induced distortions but rather to correct any systematic alignment changes to the optical system. 

\subsection{Stability \& Trending}

\begin{figure*}
  \centering
  \includegraphics[width=\textwidth]{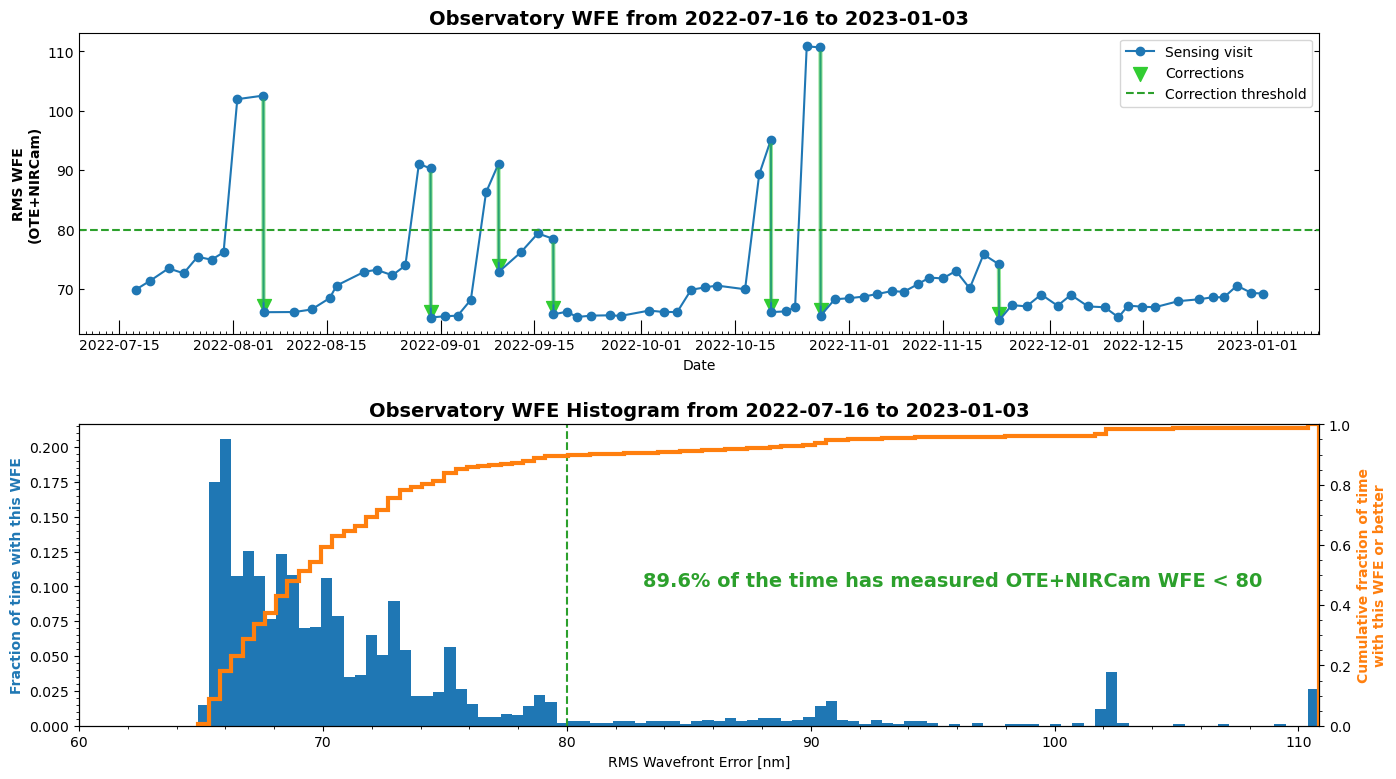}
  \caption{Top: Observatory-level WFE as a function of time since the beginning of the science mission (mid July 2022), showing every sensing visit along with the few instances where corrections to the PMSAs were applied. Occasional larger tilt events are responsible for the larger the wavefront changes over time, as discussed in the text. Bottom: histogram of the WFE so far showing that a large majority of observations have near-optimal image quality. A small fraction of the time has WFE above the correction threshold, generally the time between a larger tilt event and its subsequent correction. PSFs during such times can have modest but detectable differences from the average PSF, which can be calibrated during science analyses using the wavefront data. The observatory-level WFE mission requirements have so far been met at all times within this period.} 
  \label{fig:WavefrontTrending}
\end{figure*}

The baseline operations concept outlined above has been operational since mid-July 2022 and trending of the WFE, a key performance metric, along with the occasional corrections that were made to the PMSA are shown in Figure~\ref{fig:WavefrontTrending}. Since the start of the science mission, 47 wavefront sensing observations have been executed, including 6 that included PSMA control. This cadence averages out to correcting once every 2.6 weeks, i.e. better than our expectations. In general, the OTE has been stable to within $\sim$10~nm RMS until a disturbance occurs and must be corrected. Figure~\ref{fig:WavefrontTrending} suggests that corrections are grouped closer together, separated by longer periods of stable WFE. Most of the sudden changes in WFE seen in Figure~\ref{fig:WavefrontTrending} are referred to as tilt events and are discussed below in Section~\ref{subsubsec:TiltEvents}.

Overall, the telescope's performance has met the criteria listed above about 84$\%$ of the time, and has met the mission requirements 100$\%$ of the time since mid-July 2022.

% Wavefront Trending

\subsubsection{Tilt Events} \label{subsubsec:TiltEvents}

\begin{figure*}
  \centering
  \includegraphics[width=0.9\textwidth]{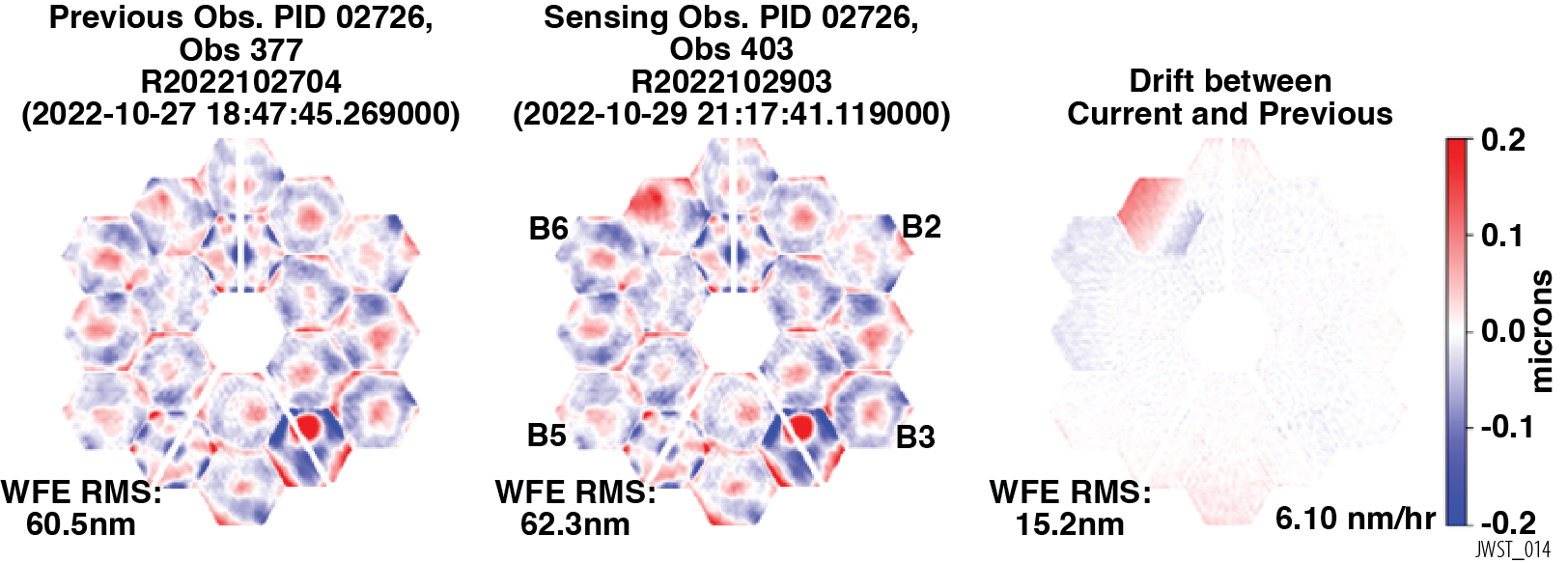}
  \caption{Example of a single-segment tilt event, as seen in optical path difference maps measured during wavefront sensing observations. Only one segment, C6, moved much in this recent event; other events have shown correlated motions of several segments, apparently related to release of tension within the wing hinge areas. When such events lead to the total WFE surpassing the correction threshold, corrective mirror moves are scheduled for the subsequent WFS observation.  Segments not affected by any tilt event generally show superb stability, often below the 7~nm sensing noise, as seen above for the right half of the primary.
  } 
  \label{fig:TiltEvent}
\end{figure*}

On many occasions, so-called tilt events, where sudden and uncommanded tilts of individual or groups of segments (e.g. wing segments), have been observed throughout commissioning and science operations. These tilt events were first observed during OTIS cryo-vacuum testing at Johnson Space Center in 2017 and they have been ascribed to the stick/slip strain release stored in the OTE hardware and/or structure during cooldown. They are expected to decrease in numbers over time as the OTE structure and hardware relax into their new environment. Tilt events continue to occur in the science mission (e.g., \citealt{Schlawin+Beatty+Brooks+etal_2022}, in preparation), though less frequently and at a lesser magnitude than during early commissioning. As shown in Figure~\ref{fig:WavefrontTrending}, tilt events episodically punctuate weeks-long periods of wavefront stability. In practice, the infrequent occurrences of large tilt events have been the dominant source of WFE degradation requiring PMSA correction. Note however, that not all tilt events have led to a PMSA correction, and those who did were all corrected as part of our routine maintenance program. Ongoing trending will track the nature and frequency of these events. The last two months of 2023 did not have any tilt events that drove excursions to 80~nm control threshold, which supports the hypothesis that the OTE structure is relaxing to a stable state. An example of a tilt event that occurred between wavefront sensing visits is shown in Figure~\ref{fig:TiltEvent}.

% Tilt Events

\subsubsection{Micrometeoroids} \label{subsubsec:micrometeoroids}

\begin{figure}
  \centering
  \includegraphics[width=0.45\textwidth]{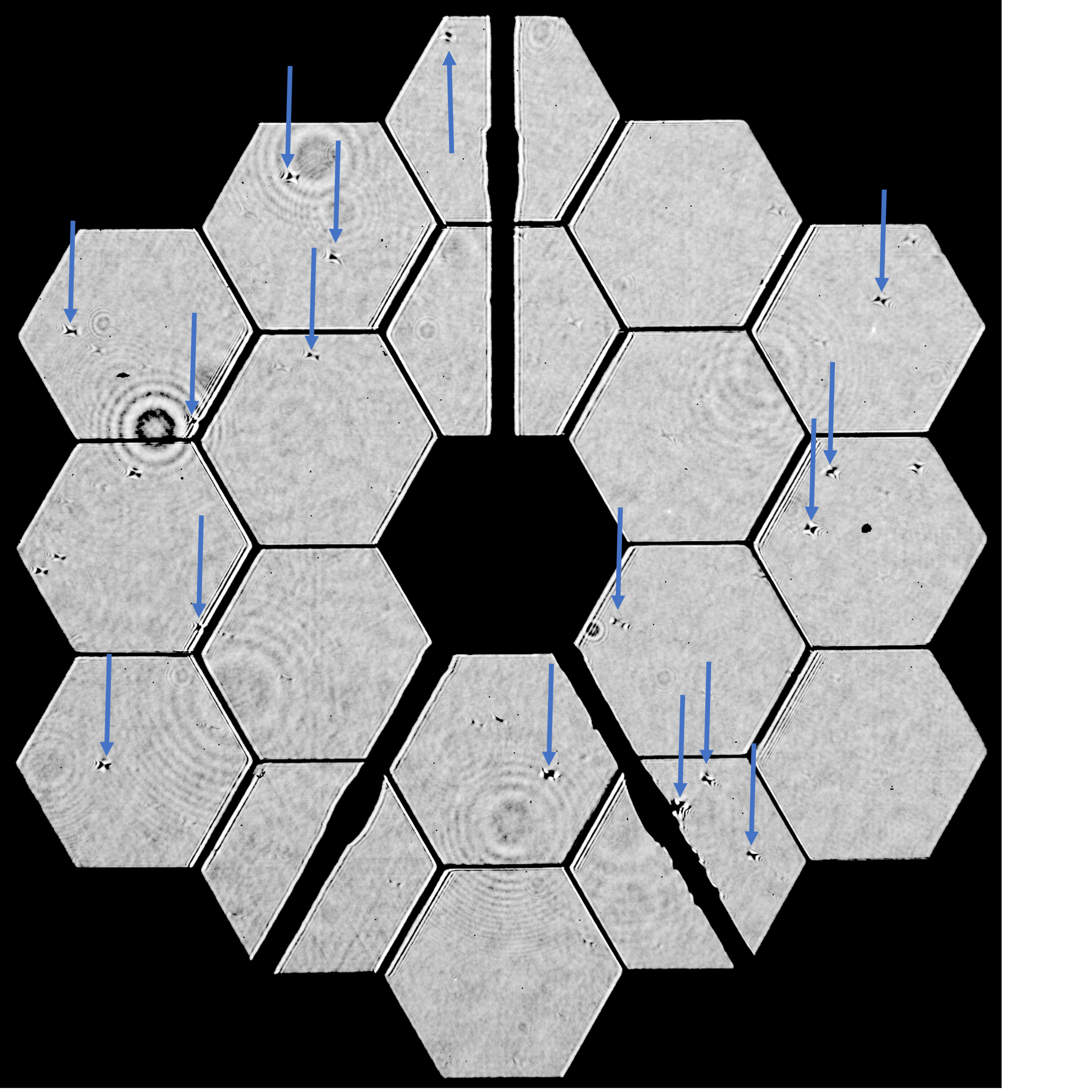}
  \caption{The NIRCam pupil imaging lens is used to monitor the telescope mirrors on a quarterly basis. Localized disturbances to the wavefront due to micrometeoroid impacts are visible in the logarithmically stretched image above. Some of the micrometeoroid events are detected by wavefront changes during routine sensing observations, and the corresponding pupil image features are marked here with the blue arrows. The large C3 segment event, which had the largest wavefront effect, can be seen next to the secondary mirror support strut (bottom right). The other features include pre-flight contaminants along the optical path, most of which have been stable throughout the ground test program. The appearance of uneven segment gaps is due to pupil image aberration and diffraction.}
  \label{fig:Micrometeoroids}
\end{figure}

Impacts from micrometeoroids on the PMSA have been observed since the middle of OTE commissioning. Pupil imaging first revealed localized surface changes to individual mirror segments, and phase retrieval analysis has revealed, in some cases, WFE changes on the impacted segments. However, not all micrometeoroid impacts have resulted in measurable changes in WFE since some show up only on pupil images or when averaging large numbers of optical path difference (OPD) maps (Figure~\ref{fig:Micrometeoroids}). Moreover, the cumulative effect of these micrometeoroids impacts has so far minimally affected the overall telescope throughput. 

Notably, however, a large impact on segment C3 was observed in phase retrieval analysis from sensing visits covering the period 22-24 May UT. The impact was such that the global WFE worsened by 9~nm RMS, after compensation by applying segment corrections in all degrees of freedom. 

The telescope wavefront error is still well below the nominal requirement following the single C3-event.  However models of similar events indicate that with about ten similar events, we could be at our end-of-life wavefront error requirement of 150~nm RMS. Due to the precision launch of JWST, the observatory has sufficient fuel for 20+ years of mission life, considerably longer than the mission design requirement minimum of 5 years. With the possibility of an extended mission and the uncertain rate of C3-type events (from only a single occurrence) and the unexpected resulting WFE, the project has implemented a meteoroid avoidance zone (MAZ) for Cycle 2. Models produced by NASA's Meteoroid Environment Office show that the greatest impact rate for higher energy micrometeoroid strikes from sporadic sources, occurs in the so-called ram direction, the direction of flight as JWST moves with the Earth around the Sun. The Cycle 2 pointing restrictions favor observing in the wake (anti-ram) direction whenever possible. The proposed MAZ would reduce the instantaneous field of regard by about 40$\%$. Models which remove all pointings from the MAZ and redistribute them over the allowed field of regard, can lower the impact rate on the primary mirror by 55-65$\%$. The reduction we expect to achieve in practice will likely be 30-40\%, because some high-priority and time-critical pointings in the MAZ will be allowed. As the Cycle 2 detailed observing plan is constructed, the expected impact reduction rate will be determined and monitored throughout the Cycle.  Detailed damage models are also being constructed to provide a better understanding of the true risk of further C3-events; preliminary models suggest this was a higher than average energy impact on a sensitive area of C3. These models will be supplemented by data from a series of ballistic tests on relevant samples which should be completed over the next few months, building upon the experimental testing carried out early in the JWST development (e.g.,  \citealt{Heaney+Pearl+Stuebig+etal_2004}).

The impact rate appears consistent with pre-flight expectations and mitigation strategies are being implemented. Meanwhile, micrometeoroids are being partially corrected as part of our routine maintenance corrections using, in particular, radius of curvature actuators. Since the end of commissioning, we have gained experience over a longer time baseline and with larger-number statistics.

% Micrometeoroids

\subsection{PSF Calibrations for Science Observations}

Every wavefront sensing and control observation produces, by means of phase retrieval analysis, optical path difference maps that can be used to model the point spread function. These maps are publicly available (see https://mast.stsci.edu/) and can be readily imported using the WebbPSF Python package \citep{Perrin+Sivaramakrishnan+Lajoie+etal_2014}, which now provides the capability to generate model PSFs that track the measured variations in mirror alignment over time. Efforts continue to further refine and improve PSF models based on in-flight experience and detailed comparisons with flight data. The package also includes trending features, some of which were used to generate Figure~\ref{fig:WavefrontTrending}. Though science observations inevitably encounter PSF variations over time, sometimes more than others, for many science use cases the impact of such variations can be reduced as long as those variations are measured and characterized. 

\section{Discussion and Lessons Learned} \label{sec:LessonsLearned}

JWST was a first-of-its-kind large segmented cryogenic space telescope.  This unique telescope brought with it many challenges, requiring the team to investigate how to mount and align a segmented mirror, how to test a large cryogenic telescope, and how to commission this complicated system. In the end, this highly complex, large, cryogenic, deployable space optical system has worked extremely well. The deployments and telescope alignment activities were executed smoothly, without incidents or invoking any contingencies. Several minor surprises were all handled within the normal course of events as part of the commissioning process and without any particular difficulty. The optical performance budgets were all met or exceeded, enabled by following the systems engineering framework. The telescope hardware, wavefront sensing and control algorithms, systems, processes, and our dedicated teams all worked as planned. 
Future telescopes can benefit from and build upon these lessons learned; some have already been published during the development period (e.g., \citealt{Stahl_2010}, \citealt{Feinberg+Arenberg+Yanatsis+etal_2018}), while others are still being formulated and documented. It is an appropriate time to provide some discussion as to what led to this success and what could be done better next time. 

\subsection{Space-Based Deployable Optics and Active Wavefront Control Flight Proven}

It is now obvious, but still noteworthy, that the complex sequence of events involved in deploying and aligning the OTE was successful. Given the unprecedented nature of this mission, this was not assured to be the case ahead of time. Achieving this required state-of-the-art contributions across a vast range of engineering disciplines from countless individuals and many organizations. The successes in flight are a testament to the efficacy and rigor of these long engineering processes.

JWST's wavefront sensing achieved extraordinary precision. The wavefront sensing error budget included several terms that were not possible to test to high precision and the actual performance worked in our favor. For example, the fast wavefront sensing implemented for the IEC and thermal stability tests demonstrated sub-nm fast differential measurements, far better than expected. The NIRCam-only multi field sensing measurement (coarse-MIMF) fully corrected the multi-field alignment, which was confirmed when no further corrections were warranted after multi-field sensing measurements. As yet another example, in the final iteration of the coarse phasing activity, the piston errors were so small that our methods struggled to sense them.

The mirror control benefited from actuator moves that were more precise than budgeted. This made the overall process more efficient and effective, with the ultimate achieved alignment that is significantly better than the requirements. The level of conservatism in the error budgeting and model uncertainty factors was appropriate engineering given the first of a kind nature of this process, but it is now clear how well this all worked.

\subsection{Piecewise Verification by Analysis Works}

The driving telescope performance parameters could not be directly tested and therefore needed to be verified by analysis. This was unprecedented in scope and required substantial development of high-fidelity integrated models. The integrated modeling effort naturally grew in scale and fidelity as the design matured, with initial efforts focused on design trades and ultimately converging to a very accurate end-to-end model used to verify performance for the pre-ship review. The model size was complex as it required predictive accuracy on nanometer scales for structures that were meters in size and used novel materials operating over a wide range of temperatures. These new models were managed with rigorous oversight for model construction, verification, and validation at each level of integration. The team intentionally adopted what were thought to be conservative model uncertainty factors, which were in some cases found to be just right for bounding flight performance, and in others overly conservative. 

The JSC cryotest was complex. Nevertheless it proved its value, not just in validation of the overall performance and requirement verification, but also by allowing the discovery of several workmanship issues (particularly with the integration of the frill) that could be addressed prior to launch. 

Piecewise verification by analysis, combined with adequate testing to validate workmanship, was critical to the success of JWST.

\subsection{Application of Lessons from Hubble and Chandra} \label{subsec:LessonsHubbleChandra}

The JWST telescope implemented important lessons learned from the Hubble and \textit{Chandra X-Ray Observatory} (Chandra) programs and these were considered critical to the success of this mission. The integration and testing approach was developed in a way that considered the Hubble program \citep{Feinberg+Geithner_2008}. For JWST, there were pre-defined test criteria established and a process for reporting if those tests are not met. In the case of JWST, formal processes were defined for problem reports, anomalies, and failures. JWST also developed a matrix of crosschecks, both in measurement and in analysis, that was used to catch errors and provide more confidence in the results. The results were transparently reported at data reviews and evaluated by the project team, external review boards, and external advisory groups such as the optical PIT (see Section~\ref{sec:TelescopeDevelopment}). 

We have previously noted the importance of independent test equipment; here we highlight the importance of independent analyses as well. JWST designated a walled-off group within the project team as an Independent Optical Verification Team. This group was charged with conducting independent analyses of key optical results, using independently developed software tools and often different analysis approaches from the primary optical verification team. This independent check was critical in ensuring the validity of test results, as well as of the numerous optical models and test configuration metrology that underpinned those test results.

The JWST Pathfinder Program (see Section~\ref{subsubsec:Pathfinder}) was a hallmark of the JWST I\&T program that provided invaluable risk reductions while proving out new systems for handling, integration, and testing \citep{Feinberg+Keski-kuha+Atkinson+etal_2010}. The Pathfinder Program was derived from a successful verification engineering test article program on Chandra \citep{Arenberg+Matthews+Atkinson+etal_2014}. Although there were significant costs incurred on the Pathfinder Program, it matured all of the processes needed. This ultimately streamlined many activities by allowing the evaluation of test environments prior to flight tests and avoiding incidents on the flight hardware itself; the OTIS cryotest met all of its test objectives the first time without incident and without delays. 

The Pathfinder program was particularly valuable for Webb given the heightened challenges of the mission. Cryo-vacuum testing always presents challenges, but JWST's size, wide range of payload operating temperatures, and infrared wavelength coverage presented exceptional demands regarding lengthy thermal transitions, contamination control, control of thermal background light, macroscopic motions and interferences during cooldown, jitter control, and careful management of atmospheric condensation onto chamber shrouds. The details of how these issues were managed for JWST are outside the scope of this paper, but discussions of such topics can be found in published overviews of the major ISIM and OTIS cryo-vacuum tests \citep{Kimble+Vila_VanCampen+etal_2016, Kimble+Feinberg+Voyton+etal_2018}.

\subsection{Balancing Development Risks and Science Performance} \label{subsec:VerSciPerformance}

Mission requirements are defined and rationalized through a science traceability matrix, which then drives mission level requirements and lower-level requirements. However, this science-driven approach can sometimes lead to system-level performances that are difficult to achieve, resulting in considerable schedule and cost growth. For example, in 2005 the JWST Science Assessment Team recommended, and the Science Working Group approved, relaxation of strict contamination requirements in favor of mirror cleaning procedures \citep{JWST_SAT_Report}. These procedures helped keep contamination budgets at manageable levels and reduce cost. Future missions should set realistic contamination levels at the outset, which necessitates early discussion of I\&T.

The same review eliminated the 1~$\mu$m encircled energy requirement \citep{JWST_SAT_Report}. There were concerns with verifying the mirror performance at 1~$\mu$m given factors such as the convergence rate to meet the polishing specification, creep and micro yield mirror changes, and individual deformation on mirrors from backplane deformation. The decision was to tighten the low-frequency wavefront error allocation while increasing the mid- and high-frequency allocations. This lead to significant risk mitigation by providing mirror polishing schedule relief, relaxation of challenging error budget terms, and relaxation of alignment tolerances. As reported in Section~\ref{subsubsec:ObsWFE}, the image quality observed at the end of commissioning was diffraction-limited at 1.1~$\mu$m in spite of this requirement relaxation. 

\subsection{Stability of a Large Space Telescope}

The telescope architecture relied on the support structure to provide passive stability with wavefront sensing every other day and control no more than every two weeks. This wavefront sensing and control operations concept has proven to be appropriate as demonstrated in Figure~\ref{fig:WavefrontTrending}. 

One of the larger uncertainties pre-launch was whether observatory stability levels would meet the predictions from the integrated modeling. The observed drift sources and amplitudes were anticipated and accurately modeled, as described in Table~\ref{tab:StabilityPerformance}. However, the thermal transient models were not validated and for reasons not yet understood, some transient thermal drifts were significantly over-predicted, and the thermal distortion time constants came out significantly shorter than predicted.

Although the IEC heaters’ on-off cycling with bang-bang control results in a wavefront oscillation that is quite small (only $\sim$2.5~nm), the effect is nonetheless easily sensed in high-precision transit observations. This IEC heater instability was due to the cable harness connection to the telescope being very stiff, which was not initially captured in the telescope thermal distortion integrated model. The integrated models did predict the observed level of drifts with remarkable accuracy when their properties were included (see Table~\ref{tab:StabilityPerformance}). Modeling needs to carefully consider the harnesses, and such bang-bang control heaters should be used with caution in future missions striving for ultra stability.

Tilt events have continued into the science mission, which is not entirely unexpected, but they are still the dominant source of wavefront changes. Tilt events are easily sensed and corrected by the wavefront control system. The expectation is that the spontaneous tilt events will continue to subside as the structure relieves stress, but continued monitoring will help determine the nature of these events. JWST material and cryogenic testing showed that the tilt events were infrequent and sufficiently small to satisfy JWST's encircled energy stability requirements. For a future system that requires picometer stability, it will be important to demonstrate that picometer-class lurches are understood early (in the technology phase) such that the active control systems can be designed to sense and correct these displacements.

\subsection{Team Cohesiveness and Rehearsals}

The importance of an effective, well-trained, integrated and ``badgeless'' team is well known, but worth repeating. Several factors contributed to the development of such a team for the JWST. The multi-institutional team comprised members from the government, academia, and the aerospace industry spread across the United States. The telescope management team shared clear goals and created an environment that encouraged cross-team interactions, communications, and information sharing. The management team worked hard to foster trust and engender transparency across the distributed team, which we consider critical components of the mission success. The regular interactions and open communications allowed the team to work effectively together, identify problems and formulate solutions. 

The team was confident and well-prepared to execute the mission, drawing on experience gained from the test program and many dedicated commissioning rehearsals. These rehearsals were operationally flight-like, incorporating procedures and high-fidelity simulations for data analysis. Comprehensive, detailed operational processes and procedures, critical to a mission of this complexity, were improved over time given the lessons learned from rehearsals. These rigorous rehearsals enabled the team to prepare and become comfortable with all aspects of the commissioning process prior to launch. 

The team also benefited from an unusually high degree of continuity of staff on JWST, including key members in leadership positions. This continuity formed a base of multi-institutional knowledge and long relationships that proved valuable in problem solving. Many had experience with much of JWST mission's life, with some being involved from its inception through the end of commissioning. The team also benefited from significant experience with other space telescope projects, such as HST.

\subsection{Guiding and Line of Sight Performance} \label{subsec:GuidingLOS-LessonLearned}

JWST has demonstrated stable pointing with extremely low jitter. JWST routinely achieves LOS jitter $\sim$1~mas RMS through the use of a fine steering mirror in the control loop, and via high cadence subarray readouts using a focal plane instrument at pixel scales equivalent to the science instruments. For comparison, Hubble's RMS jitter is typically $\sim$2.5 to 3 mas \citep{Lallo_2012}. Observational techniques to obtain and analyze the jitter data were proven to be successful \citep{Hartig_2022}, and results show the JWST design effectively isolated vibration sources such as the cryocooler and suppressed other potential contributors to jitter.

Complex attitude control systems are traditionally challenging to fully test pre-flight, and JWST's was no exception. Even though a closed loop guiding demonstration was executed during  OTIS CV tests (Section~\ref{subsubsec:OTIS-cryovac}), it was necessarily lacking in full fidelity. Furthermore, early OTE commissioning activities required close-loop guiding with a misaligned telescope, when the transformations from ACS to the focal plane had not yet been established and guidestar PSFs were not yet stacked. (Sections~\ref{subsubsec:GuidingCommissioning} and~\ref{subsubsec:CalTel2Spacecraft})

From early in the commissioning planning process, it was recognized that this called for particularly close coordination among the teams responsible for Attitude Control, Wavefront, and ISIM/FGS in developing the commissioning activities. The teams worked through the details of non-standard guiding scenarios with control loop components that were not yet fully calibrated, developed the operations concepts for manually updating ACS transformations, produced the tools for overriding the nominal guiding where necessary, and wrote a number of contingency plans. As a result, in flight, the sometimes subtle interdependencies of the ACS, FGS, and the OTE optical alignment process came as little surprise, and the team was able to effectively navigate along the road to a successful and complete OTE commissioning.

The generalized lesson from this experience is that a spacecraft's pointing and attitude control system is a key component to its science performance, and benefits from being treated as such, holistically, from early in the commissioning plans. Yet, the integration and consideration of the ACS as a system fundamental to science commissioning can be complicated by cultural differences in approach, language, and tools. This was anticipated from prior mission experience and mitigated by an integrated inclusive approach to OTE commissioning in general. 

\subsection{Importance of Determining Test Configurations Early}

While a high-level summary of the I\&T program is provided in Section~\ref{subsec:IntTest}, the test activities changed dramatically in scope and implementation during the development phase. An earlier definition of test configurations could have saved resources. For example, the initial telescope cryotest concept had the telescope pointing down, called ``cup down'', on a $\sim$300,000 kg stainless steel tower with six spinning and rotating cryogenic autocollimating flats and cones of light through instruments to measure the alignment by stitching interferograms. The extreme complexity of the cup down test led to a total redesign that resulted in a simplified, yet still challenging to execute, ``cup up" configuration as described in Section~\ref{subsubsec:OTIS-cryovac} \citep{Atkinson+Arenberg+Waldman+etal_2008}. The cup down configuration was originally baselined in order to prevent any contamination of the telescope mirrors, but the contamination was later deemed manageable in the cup up configuration by implementing a mirror-cleaning procedure. This mirror cleaning procedure was used following the OTIS cryotest and before shipping to the launch site \citep{Lobmeyer+Carey_2018,Abeel+Huang_2018}.

\subsection{Micrometeoroid Environment and Damage}

The open architecture of the JWST telescope makes the primary and secondary mirror optics particularly susceptible to micrometeoroid damage. As discussed in Section~\ref{subsubsec:micrometeoroids}, the effect from micrometeoroids was estimated based on the environment and a damage model from impacts. The micrometeor effects may have been underestimated, though the bulk of the damage to date has come from the single event on C3, so there is great statistical uncertainty in how the damage rate should be extrapolated into the future. The impact physics is also complex and uncertain; ground tests can't achieve micrometeoroid-like velocities and neglected how cryogenic temperatures change the material stiffness properties. Because of the statistical and damage model uncertainties, and with the prospect for a JWST mission lifetime far exceeding requirements (due to available propellant), restrictions on the field of regard will be implemented to reduce ram direction micrometeoroid impact rates. Missions under consideration with even larger aperture optics and even tighter wavefront requirements will need to consider this issue carefully and track JWST's experience as it accumulates a longer baseline of operations.

\subsection{Optical Modeling Across Interfaces} \label{sec:opticalmodeling-lesson}

In the development of the independent system elements, there were separate optical and optomechanical models for the telescope and each of the science instruments via the interface shown in Figure~\ref{fig:OpticalDesign}. Early in the program as the prime contractor and the NIRCam were selected, it was realized that the optical designs were not compatible. To deal with this, the telescope design was modified and the F-number changed to accommodate the interface. However, it was later uncovered that this design change increased susceptibility to the rogue path that was described briefly in Section~\ref{subsec:DesignImplementation} and more thoroughly in \cite{Lightsey+Wei+Skelton+etal_2014}. The rogue path passes through the AOS entrance aperture and directly onto the science instrument pick off mirrors, where it then enters the science instrument optical path through reflections or scattering. 

The rogue path stray light was well known and modeled. The models were used to confirm no direct optical paths to the detector focal planes and to determine scattering from the pick off mirrors was negligible. However, what was missed, was the possibility of rogue path stray light causing grazing angle scattering off instrument structure from the pickoff mirror housing and downstream in the optical path of the instruments. After observing the stray light in flight (see Section~5.3 of Rigby PASP Science), the optical and optomechanical models were used to reproduce the observed phenomena, confirming the stray light paths. The area of susceptibility for the observed features that was observed is a small subset of the total rogue path region of the sky. Now that the grazing angle stray light paths are known, observation scheduling can largely prevent placing bright targets in the susceptible region on the sky relative to the science target. 

Not all of the full structural as well as optical characteristics were captured in the pre-flight modeling. Programmatic constraints impeded clear communication and modeling of the full system interface. The optical prescription of the instruments was in the full system model, but not the detailed structures that included the housing around the pick-off mirrors and the detailed optomechanical structures in the science instruments. The full up system model only included NIRCam and MIRI, in order to verify the stray light light requirements at the NIR and MIR wavelengths, and the instrument teams were left to carry out the detailed analysis of their instruments. The lesson learned is that additional modeling of the complete integrated system for all modes is needed to reduce risk of unexpected stray light phenomena. This includes accurate details of mechanical structures as well as the optical prescriptions throughout the complete system.

\section{Conclusion} \label{sec:conclusion}

The revolutionary JWST telescope is performing better than all of its design objectives, enabling even higher sensitivity and more stable observations than originally planned. The telescope was made possible through the advancement of several new technologies, all of which were developed and flight proven to work as intended. The telescope has now embarked on its Cycle 1 science observations, beginning a scientific journey that will answer some of the biggest questions in astrophysics and planetary science. 

\acknowledgments

We are grateful to the JWST photography team, especially Chris Gunn, for capturing the JWST development, and Heather Ghannadian for graphics design.

The JWST mission is a joint project between the National Aeronautics and Space Agency, European Space Agency, and the Canadian Space Agency. The JWST telescope development was led at NASA's Goddard Space Flight Center with a distributed team across Northrop Grumman Corporation, Ball Aerospace, L3Harris Technologies, the Space Telescope Science Institute, and many other companies and institutions. This telescope was created by a large team of people from many diverse backgrounds whose creativity, passion, teamwork and endless sacrifices made this scientific dream a reality.

\bibliography{biblio.bib}

\end{document}